\documentclass[letterpaper]{article} 
\usepackage[draft]{aaai2026}  
\usepackage{times}  
\usepackage{helvet}  
\usepackage{courier}  
\usepackage[hyphens]{url}  
\usepackage{graphicx} 
\usepackage{natbib}  
\usepackage{caption} 
\usepackage{xcolor}
\usepackage[toc,page]{appendix}
\usepackage[english]{babel}
\newcommand{\answerYes}[1]{\textcolor{blue}{#1}} 
\newcommand{\answerNo}[1]{\textcolor{teal}{#1}} 
\newcommand{\answerNA}[1]{\textcolor{gray}{#1}}

\definecolor{color0}{RGB}{0,0,0}

\newif\ifrevmarks
\revmarksfalse
\ifrevmarks
  \definecolor{revcolor}{RGB}{117,39,163}
\else
  \colorlet{revcolor}{black}
\fi
\newif\ifchecklist
\checklistfalse
\newcommand{\rev}[1]{\textcolor{revcolor}{#1}}

\usepackage{array} 
\newcolumntype{Y}{>{\raggedright\arraybackslash\let\\\tabularnewline}X}

\usepackage{amsmath}
\usepackage{graphicx}
\usepackage{subcaption}
\usepackage{tikz}
\usetikzlibrary{shapes.geometric, arrows.meta, positioning, fit, backgrounds, calc, decorations.pathreplacing, matrix}
\usepackage{booktabs}
\usepackage{tabularx}
\usepackage{booktabs}

\title{Mapping the Winds of Stance Dynamics using Potential Landscape Models}
\author{
      Benjamin Steel\textsuperscript{\rm 1},
      Derek Ruths\textsuperscript{\rm 1}
  }
  \affiliations{
      \textsuperscript{\rm 1}McGill University \\
      \texttt{benjamin.steel@mail.mcgill.ca},
      \texttt{derek.ruths@mcgill.ca}
  }

\begin{document}

\maketitle

\begin{abstract}
    From changing fashion trends to views on world leaders and economic policies, large-scale shifts in group positions happen regularly and unexpectedly. How can we track these in the wild? How can we characterize them? Existing work has primarily leveraged stance detection to track shifts of specific groups on a single issue. However, such methods will only find shifts when they accurately pick exactly the right group and right issue. They do not capture the multi-dimensional, multi-resolution stance landscape in which these shifts actually happen. To better model drift and shift in public opinion, we require a framework that can track change at the population level, across a diverse range of issues. We propose a method to infer the potential landscape of stance dynamics, the gradient of which shows large-scale stance shifts, and apply it to show \rev{en masse} stance shifts by prominent Canadian political \rev{elites} across multiple platforms and years. We do this using large-scale stance detection to find stance expressions, \rev{latent-GP factor analysis,} and potential landscape neural networks to find the potential landscape of that space. This allows us to find \rev{descriptive potential landscapes in} a coherent, linear, \rev{low-dimensional space,} where we can explain the specific characteristics of each dimension. \rev{As confirmation of the descriptive power of the model, it also has predictive power: it is significantly better than a simple baseline, and beats an auto-regressive forecasting method at short horizons, and a damped-Theta forecasting method at long horizons.}

\end{abstract}

\begin{links}
    \link{Code}{https://github.com/bendavidsteel/latent-stance-dynamics/}
\end{links}

\section{Introduction}
\label{sec:intro}

\begin{figure}
    \centering
    \includegraphics[width=0.65\linewidth]{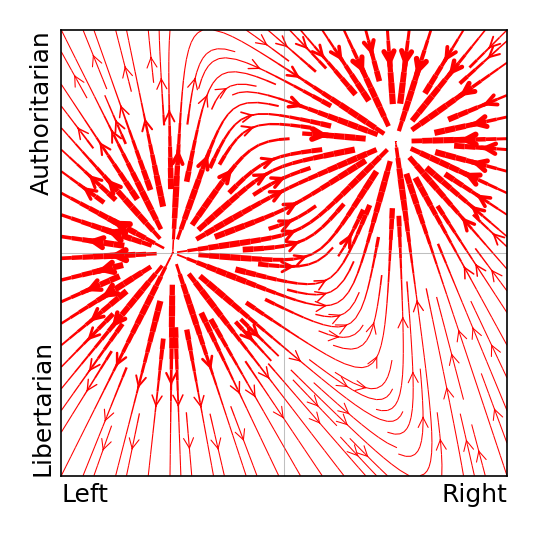}
    \caption{\textit{`This year, people on the left have diverged in their stated preference for authoritarian or libertarian systems, but people on the right have converged in support for authoritarianism'} \textemdash how can we re-construct this fictional, idealized now-cast in a data-driven, empirical fashion?}
    \label{fig:explanatory}
\end{figure}

How can we map when people change their minds en masse? Can we do so in a descriptive and predictive fashion? Can we find a sufficiently low-dimensional space that allows us to simply characterize the shifts happening? A method that can produce a map such as shown in Fig. \ref{fig:explanatory} would provide powerful empirical insight into how stances change over time.

We propose a method to produce a map of stance dynamics, which infers the potential landscape that shows when and where there is movement across this \rev{space,} and validate this model \rev{by measuring its predictive power.} We apply this method to social media data here, meaning the latent space we find is of stances expressed online. However, this method could be used for any text data, for example transcripts from public deliberation, or high dimensional longitudinal survey data.  We infer latent stance trajectories from a \rev{multi-year} cross-platform dataset of prominent members of Canadian political \rev{society, who are the main agenda setters in Canadian political discourse,} and use them to find the density of people \rev{in different areas and periods} in stance space. Then, to find common movement of stance trajectories in this space, we infer a uniform-field, in \rev{the} form of a potential landscape of that space, the gradient of which tells us the prevailing direction of movement at any point.

Our work uses ideas, methods, and inspiration from biophysics by considering movement of individuals in stance space as a Langevin process \cite{uhlenbeck1930theory}, following work from \citet{achitouv2026d} and \citet{iniguez2012opinions}. This formulation models individual stance movement as following the gradient of a potential landscape, and with all other movement modelled as noise. Our application, in which individual entities both create and are acted on by a potential landscape in the absence of an explicit physical medium, is known as an active matter system in physics, and has been used to model fish schooling and birds flocking \cite{marchetti2013hydrodynamics}, showing us that Langevin process models can be successfully applied in these areas. In our application, we think of this potential landscape as encoding the effect of news \citep{sevenans2018mass,arceneaux2018elites}, trends, and elite cues \citep{broockman2017causal}, which affect mass publics and elites alike. People at different attitudinal positions and different times in this landscape are impacted by new information differently. While the simple form of the potential landscape we present here may not represent the true complexity of how information affects stances, this is a reasonable initial simplification. Though we theorize the potential landscape has a causal effect on stances, here we use it as a descriptive tool to find collective stance movement.

This takes a different perspective from much work in opinion dynamics, where individual movement in stance space is modelled as entirely or partially determined by interactions with other individuals, as in agent-based models \citep{peralta2022opinion,okawa2022predicting,monti2020learning}. While stances likely change due to both uniform-field effects and individual interaction-based influence, and myriad more-complex mechanisms, the uniform-field effect has been under-explored \citep{achitouv2026d,introne2023measuring,iniguez2012opinions} compared to interaction-effect models in computational social-science. The uniform-field has the benefit of also encoding some information that may be considered an interaction effect: influence from co-partisans, party elites, and audience reactions coming via shared channels.

We therefore instead focus on finding large-scale stance movement. We know that groups of people frequently change their attitudes together in response to information in a process called the `bandwagon effect' \citep{farjam2024social}, or do not voice dissenting attitudes in a `spiral of silence' \citep{noelle1974spiral}. Similarly, politicians are known to follow the cues of prominent partisans of their party \citep{barbera2019leads}. Conversely, voters are known to follow the lead of politicians \citep{broockman2017causal}. People perceive these attitude changes through their local information environment \textemdash those who are in their political in-group, an effect exacerbated online. We design a method to find those aggregate movements, places where people change their stances together due to social influence. We will use a model that we optimize by predicting future stances, but where the model is explicitly formulated to produce a visualizable landscape of stance movement, validating descriptive-ness via predictive-ness.

This work extends the application of stance detection, the task of inferring the attitude of the author of a piece of media \citep{mohammad2016semeval,bestvater2023sentiment}. Previous work has frequently selected specific topics or entities, known as stance targets in the field of stance detection, to find stance shifts on. This work broadens this to use methods that discover stance targets automatically. This allows us to maximize the number of stance expressions for a given dataset size, more robustly understand stances by covering many specific targets \citep{druckman2019we}, and find instances of stance change when it is not on the same specific stance target.

Although the stances expressed on social media are not the same as the private attitudes expressed in surveys \citep{kuran1998private,noelle1974spiral}, this work uses a dataset of posts from prominent public figures. As such, whether or not they believe their own stance expressions, they are consumed by the population at large, and as such are some of the most important to understand regardless.

In general, we seek to answer the following research question: \rev{\textbf{Can we find a descriptive and explainable model of stance dynamics on social media, which is validated by predictive power?}} Our contributions are:

\begin{itemize}
    \item We propose a novel method of finding the latent space of stance from media content.
    \item We apply dynamical methods to stance trajectories to determine their statistical properties, and show general stance movement in this latent space.
    \item We characterize the \rev{predictive power} of the potential landscape model, showing that \rev{it is significantly better than a simple baseline, and beats time-series forecasting methods on some horizons.}
    \item We use cross-platform methods and data to widen our results beyond one platform.
\end{itemize}

\section{Background}
\label{sec:background}

\paragraph{Dynamical-Systems in Biology}

While dynamical methods were initially used in fluid dynamics, they have been increasingly applied to problems in biology. There is a long lineage of inferring potential landscapes in cell development \citep{wang2011quantifying,cislo2025reconstructing,mochulska2025generative}, and other application areas in biology \citep{grauer2021active}. \citet{howe2025learning} use a constrained neural network to parameterize the landscape from low-dimensional trajectories, using principal component analysis (PCA) to do dimensionality reduction on the high-dimensional data and learning the landscape potential. We use their method, with \rev{adaptations, including replacing the PCA with a latent-GP factor analysis model,} discussed in Sec. \ref{sec:method}.

\paragraph{Computational Studies of Attitude}

Stance detection has previously been used to measure stance expressed online, but these works only use platform-specific features \cite{villa2025social, gambini2023tweets, peng2024online, zhang2024doubleh}, or look at a small pre-defined set of stance targets \cite{ibrahim2024analyzing, sun2024exploring}. Our work looks at stance across multiple platforms, and finds stance targets from the data, both increasing the amount of stance data we can use from our dataset, and expanding beyond uni-dimensional attitude models.

Prior work in opinion dynamics developed theoretical models of attitude change due to uniform fields representing media \citep{iniguez2012opinions}, similar to our potential landscape. Other opinion dynamics work fits interaction-based stance-change models to data \cite{monti2020learning,okawa2022predicting,peralta2022opinion,starnini2025opinion,leonard2021nonlinear}, but \citet{starnini2025opinion} emphasize the need to expand this work to cross-platform data and multi-dimensional latent opinion spaces, both of which we do here. Our method does not explicitly model the effect of interaction on stance dynamics, and instead looks at general stance movement. In future, the two methods could be brought together to produce stronger models of stance change.

\citet{introne2023measuring} use fine-tuned text embeddings to construct the dynamic space of climate beliefs from Twitter data, and conduct analysis on the space via hypothesis testing. \citet{lee2025semantic} use fine-tuned text embedding models and PCA to look at the latent space of beliefs from web data, in a method similar to our own. However, they use data from a niche debate platform which is not readily generalizable to a larger population. \rev{Both} refer to the peaks of the density landscape as attractors, which is only true if the system is stationary \citep{risken1989fokker}, an untested assumption in both works. \citet{achitouv2026d} model opinion dynamics using a potential landscape model, looking at the dynamics of the pro-climate vs. denialism attitude dimension on Twitter. These papers use data from a single platform, and use \rev{embedding} models or non-linear dimensionality reduction techniques which heavily limit their interpretability, and produce artifacts that are hard to \rev{disentangle} from true dynamical landscapes. They all include an implicit and untested assumption that their system is stationary. We show that our data is not \rev{stationary here, and believe that similar studies should test the stationarity of their systems.} The potential landscape inference method used here is agnostic as to whether the system is stationary or not. We also use \rev{a} linear dimensionality reduction \rev{technique} to avoid artifacts that could obscure the true potential landscape.

\section{Method}
\label{sec:method}

\begin{figure*}[t]
\centering
\begin{tikzpicture}[
    node distance=0.5cm and 0.5cm,
    >={Stealth[length=2.5mm, width=2mm]},
    arrow/.style={->, thick, color=black!60},
    steplabel/.style={font=\scriptsize\sffamily\bfseries, text=black},
    stepbox/.style={draw=black!40, rounded corners=2pt, inner sep=3pt}
]

\node[stepbox] (step1) {
\begin{tikzpicture}[scale=0.8, transform shape,
    doc/.style={draw, rounded corners, fill=blue!10, text width=1.5cm,
                align=center, font=\tiny, inner sep=2pt},
    tgt/.style={draw, ellipse, font=\tiny, inner sep=1pt}]

    \node[doc] (doc1) at (0, 0.80) {Carbon tax now!};
    \node[doc] (doc2) at (0, -0.60) {Drop provincial tariffs};

    \node[tgt, minimum width=2.1cm, minimum height=1.0cm, fill=blue!10] (econ) at (3.0, -0.40) {};
    \node[font=\tiny, yshift=2.2mm] at (econ.south) {economic policy};
    \node[tgt, fill=blue!20] (tariffs) at (2.60, -0.28) {tariffs};
    \node[tgt, fill=blue!20] (trade)   at (3.75, -0.10) {trade};

    \node[tgt, minimum width=1.5cm, minimum height=0.7cm, fill=green!10,
          align=center] (env) at (2.90, 0.95) {climate\\action};

    \node[tgt, minimum width=1.0cm, minimum height=0.65cm, fill=yellow!30,
          fill opacity=0.75, text opacity=1] (ct) at (2.70, 0.30) {carbon tax};

    \draw[->, thin] (doc1.east) -- (ct.west);
    \draw[->, thin] (doc1.east) to[bend left=10] (env.west);
    \draw[->, thin] (doc1.east) to[bend right=6] (econ.west);
    \draw[->, thin] (doc2.east) -- (tariffs.west);
    \draw[->, thin] (doc2.east) to[bend right=10] (trade.south west);
    \draw[->, thin] (doc2.east) to[bend right=2] (econ.south west);
\end{tikzpicture}
};
\node[steplabel, above=0.0cm of step1] {1. Stance Target Extraction};

\node[stepbox, right=0.5cm of step1] (step2) {
\begin{tikzpicture}[scale=0.8, transform shape]
    \node[draw, rounded corners, fill=blue!10, minimum width=0.8cm, minimum height=0.4cm, font=\tiny, anchor=center] (post) at (-0.55, 0.8) {Post};

    \node[draw, rounded corners, fill=blue!10, minimum width=0.8cm, minimum height=0.4cm, font=\tiny, anchor=center] at (0.67, 0.92) {};
    \node[draw, rounded corners, fill=blue!10, minimum width=0.8cm, minimum height=0.4cm, font=\tiny, anchor=center] at (0.61, 0.86) {};
    \node[draw, rounded corners, fill=blue!10, minimum width=0.8cm, minimum height=0.4cm, font=\tiny, anchor=center] (target) at (0.55, 0.8) {Target};

    \coordinate (merge) at (0, 0.35);

    \node[draw, rounded corners, fill=orange!20, minimum width=1.8cm, minimum height=0.5cm, font=\tiny, anchor=center] (llm) at (0, -0.2) {LLM Classifier};

    \draw[thick] (post.south) -- (merge);
    \draw[thick] (target.south) -- (merge);
    \draw[->, thick] (merge) -- (llm.north);

    \draw[->, thick] (llm.south) -- ++(0, -0.4);

    \node[draw, rounded corners, fill=green!10, minimum height=0.3cm, font=\tiny, anchor=center] (favor) at (-1.2, -1.1) {\textcolor{green!60!black}{Favor}};
    \node[draw, rounded corners, fill=gray!10, minimum height=0.3cm, font=\tiny, anchor=center] (neutral) at (0, -1.1) {\textcolor{gray}{Neutral}};
    \node[draw, rounded corners, fill=red!10, minimum height=0.3cm, font=\tiny, anchor=center] (against) at (1.2, -1.1) {\textcolor{red!70!black}{Against}};
\end{tikzpicture}
};
\node[steplabel, above=0.0cm of step2] {2. Stance Detection};

\node[stepbox, right=0.5cm of step2] (step3) {
\begin{tikzpicture}[scale=0.65, transform shape,
    cell/.style={minimum width=0.55cm, minimum height=0.34cm, draw,
                 anchor=south west, font=\tiny, inner sep=0pt, outer sep=0pt},
    hcell/.style={cell, fill=gray!20},
    fcell/.style={cell, fill=green!20},
    ncell/.style={cell, fill=gray!10},
    acell/.style={cell, fill=red!15}
]
    \node[hcell]                        at (0.00, 1.36) {User};
    \node[hcell]                        at (0.55, 1.36) {Time};
    \node[hcell]                        at (1.10, 1.36) {Tgt};
    \node[hcell, minimum width=0.80cm]  at (1.65, 1.36) {Stance};
    \node[cell]  at (0.00, 1.02) {A};
    \node[cell]  at (0.55, 1.02) {0.4};
    \node[cell]  at (1.10, 1.02) {clim};
    \node[fcell, minimum width=0.80cm] at (1.65, 1.02) {\textcolor{green!60!black}{Favor}};

    \node[cell]  at (0.00, 0.68) {A};
    \node[cell]  at (0.55, 0.68) {0.4};
    \node[cell]  at (1.10, 0.68) {tax};
    \node[acell, minimum width=0.80cm] at (1.65, 0.68) {\textcolor{red!70!black}{Against}};

    \node[cell]  at (0.00, 0.34) {B};
    \node[cell]  at (0.55, 0.34) {1.2};
    \node[cell]  at (1.10, 0.34) {clim};
    \node[ncell, minimum width=0.80cm] at (1.65, 0.34) {\textcolor{gray}{Neutral}};

    \node[cell]  at (0.00, 0.00) {C};
    \node[cell]  at (0.55, 0.00) {2.7};
    \node[cell]  at (1.10, 0.00) {imm};
    \node[fcell, minimum width=0.80cm] at (1.65, 0.00) {\textcolor{green!60!black}{Favor}};

    \node[font=\tiny] at (1.22, -0.28) {$\vdots$};

    \draw[->, thick] (2.70, 0.70) -- (3.35, 0.70);
    \node[font=\tiny, align=center, anchor=south] at (3.02, 0.80) {latent\\GP};

    \begin{scope}[xshift=4.85cm, yshift=0.68cm]
        \draw[-{Stealth[length=1.3mm, width=1mm]}, thin]
            (-1.25,-0.88) -- (1.32,-0.88) node[right, font=\tiny] {$z_1$};
        \draw[-{Stealth[length=1.3mm, width=1mm]}, thin]
            (-1.25,-0.88) -- (-1.25,1.02) node[above, font=\tiny] {$z_2$};

        \draw[-{Stealth[scale=0.7]}, thick, blue!60]
            plot[smooth] coordinates {(-1.05,-0.30) (-1.00,0.10) (-0.90,0.45) (-0.72,0.66)};
        \foreach \p in {(-1.05,-0.30), (-1.00,0.10), (-0.90,0.45)}
            {\fill[blue!70] \p circle (1.4pt);}
        \node[font=\tiny, blue!60!black, anchor=north] at (-1.05,-0.36) {A};

        \draw[-{Stealth[scale=0.7]}, thick, red!60]
            plot[smooth] coordinates {(0.90,0.50) (0.98,0.12) (0.95,-0.27) (0.85,-0.62)};
        \foreach \p in {(0.90,0.50), (0.98,0.12), (0.95,-0.27)}
            {\fill[red!70] \p circle (1.4pt);}
        \node[font=\tiny, red!60!black, anchor=south west] at (0.94,0.54) {B};

        \draw[-{Stealth[scale=0.7]}, thick, orange!75]
            plot[smooth] coordinates {(-0.38,0.40) (-0.18,0.18) (0.02,-0.06) (0.24,-0.34)};
        \foreach \p in {(-0.38,0.40), (-0.18,0.18), (0.02,-0.06)}
            {\fill[orange!85] \p circle (1.4pt);}
        \node[font=\tiny, orange!80!black, anchor=south west] at (-0.40,0.44) {C};
        \draw[dashed, gray!60] (-0.72,0.66) ellipse (0.26cm and 0.19cm);
    \end{scope}
\end{tikzpicture}
};
\node[steplabel, above=0.0cm of step3] {3. Latent-GP Factor Analysis};

\node[stepbox, right=0.5cm of step3] (step4) {
\begin{tikzpicture}[scale=0.55, transform shape]
  \draw[-{Stealth[length=1.3mm, width=1mm]}, thin]
      (-1.90,-1.90) -- (1.98,-1.90) node[right, font=\tiny] {$z_1$};
  \draw[-{Stealth[length=1.3mm, width=1mm]}, thin]
      (-1.90,-1.90) -- (-1.90,1.98) node[above, font=\tiny] {$z_2$};
  \node[font=\tiny, anchor=north east] at (1.85,1.85) {$\phi(z,t)$};

  \draw[gray!60] (0.619,-1.021) -- (0.585,-1.012) -- (0.552,-0.994) -- (0.525,-0.972) -- (0.503,-0.945) -- (0.487,-0.916) -- (0.474,-0.883) -- (0.465,-0.843) -- (0.462,-0.803) -- (0.463,-0.763) -- (0.469,-0.723) -- (0.480,-0.683) -- (0.493,-0.652) -- (0.509,-0.621) -- (0.526,-0.594) -- (0.551,-0.565) -- (0.575,-0.541) -- (0.600,-0.522) -- (0.629,-0.504) -- (0.663,-0.488) -- (0.702,-0.474) -- (0.741,-0.466) -- (0.786,-0.462) -- (0.834,-0.464) -- (0.879,-0.473) -- (0.913,-0.485) -- (0.942,-0.501) -- (0.970,-0.523) -- (0.993,-0.549) -- (1.010,-0.581) -- (1.020,-0.615) -- (1.024,-0.656) -- (1.020,-0.696) -- (1.009,-0.736) -- (0.996,-0.770) -- (0.981,-0.799) -- (0.962,-0.830) -- (0.942,-0.857) -- (0.920,-0.883) -- (0.898,-0.906) -- (0.874,-0.929) -- (0.844,-0.952) -- (0.815,-0.972) -- (0.786,-0.988) -- (0.753,-1.003) -- (0.712,-1.016) -- (0.673,-1.023) -- (0.624,-1.022) -- cycle;   
  \draw[gray!60] (-0.839,0.465) -- (-0.879,0.473) -- (-0.913,0.485) -- (-0.947,0.504) -- (-0.972,0.524) -- (-0.993,0.549) -- (-1.010,0.581) -- (-1.021,0.616) -- (-1.024,0.661) -- (-1.019,0.701) -- (-1.009,0.736) -- (-0.995,0.772) -- (-0.979,0.803) -- (-0.962,0.830) -- (-0.942,0.857) -- (-0.918,0.886) -- (-0.894,0.910) -- (-0.870,0.932) -- (-0.844,0.952) -- (-0.815,0.972) -- (-0.782,0.990) -- (-0.751,1.004) -- (-0.712,1.016) -- (-0.673,1.023) -- (-0.619,1.021) -- (-0.585,1.012) -- (-0.552,0.994) -- (-0.526,0.973) -- (-0.503,0.945) -- (-0.487,0.916) -- (-0.474,0.883) -- (-0.466,0.848) -- (-0.462,0.808) -- (-0.463,0.763) -- (-0.469,0.723) -- (-0.479,0.687) -- (-0.492,0.653) -- (-0.509,0.621) -- (-0.526,0.594) -- (-0.548,0.567) -- (-0.571,0.545) -- (-0.600,0.522) -- (-0.629,0.504) -- (-0.663,0.488) -- (-0.697,0.476) -- (-0.741,0.466) -- (-0.786,0.462) -- (-0.834,0.464) -- cycle;   
  \draw[gray!52] (0.526,-1.217) -- (0.458,-1.200) -- (0.404,-1.172) -- (0.358,-1.132) -- (0.321,-1.086) -- (0.291,-1.034) -- (0.267,-0.974) -- (0.249,-0.910) -- (0.237,-0.843) -- (0.230,-0.768) -- (0.231,-0.687) -- (0.238,-0.616) -- (0.252,-0.547) -- (0.272,-0.487) -- (0.300,-0.429) -- (0.334,-0.380) -- (0.374,-0.339) -- (0.423,-0.303) -- (0.482,-0.274) -- (0.546,-0.252) -- (0.615,-0.238) -- (0.697,-0.230) -- (0.786,-0.231) -- (0.864,-0.240) -- (0.938,-0.256) -- (1.005,-0.278) -- (1.060,-0.305) -- (1.112,-0.340) -- (1.154,-0.380) -- (1.187,-0.429) -- (1.211,-0.492) -- (1.219,-0.563) -- (1.212,-0.638) -- (1.195,-0.701) -- (1.171,-0.759) -- (1.143,-0.814) -- (1.110,-0.865) -- (1.074,-0.914) -- (1.037,-0.959) -- (0.996,-1.002) -- (0.952,-1.043) -- (0.908,-1.079) -- (0.859,-1.115) -- (0.805,-1.148) -- (0.746,-1.177) -- (0.687,-1.199) -- (0.614,-1.216) -- (0.531,-1.217) -- cycle;   
  \draw[gray!52] (-0.815,0.233) -- (-0.888,0.244) -- (-0.957,0.261) -- (-1.020,0.285) -- (-1.075,0.314) -- (-1.123,0.349) -- (-1.164,0.394) -- (-1.196,0.447) -- (-1.214,0.509) -- (-1.219,0.585) -- (-1.208,0.656) -- (-1.188,0.719) -- (-1.163,0.776) -- (-1.133,0.830) -- (-1.101,0.879) -- (-1.065,0.927) -- (-1.025,0.972) -- (-0.986,1.012) -- (-0.941,1.052) -- (-0.893,1.091) -- (-0.844,1.124) -- (-0.790,1.156) -- (-0.732,1.183) -- (-0.668,1.205) -- (-0.590,1.218) -- (-0.507,1.214) -- (-0.443,1.194) -- (-0.389,1.161) -- (-0.346,1.119) -- (-0.311,1.070) -- (-0.283,1.017) -- (-0.262,0.959) -- (-0.245,0.892) -- (-0.234,0.821) -- (-0.229,0.745) -- (-0.232,0.665) -- (-0.242,0.594) -- (-0.257,0.530) -- (-0.279,0.469) -- (-0.308,0.416) -- (-0.345,0.368) -- (-0.389,0.327) -- (-0.438,0.295) -- (-0.497,0.268) -- (-0.565,0.247) -- (-0.639,0.235) -- (-0.722,0.229) -- (-0.810,0.233) -- cycle;   
  \draw[gray!46] (0.448,-1.364) -- (0.277,-1.297) -- (0.169,-1.174) -- (0.096,-1.017) -- (0.048,-0.843) -- (0.015,-0.652) -- (-0.012,-0.452) -- (-0.043,-0.260) -- (-0.119,-0.105) -- (-0.286,-0.037) -- (-0.497,-0.006) -- (-0.707,0.023) -- (-0.908,0.064) -- (-1.084,0.122) -- (-1.233,0.211) -- (-1.335,0.340) -- (-1.364,0.536) -- (-1.317,0.710) -- (-1.236,0.861) -- (-1.138,0.991) -- (-1.022,1.110) -- (-0.893,1.214) -- (-0.744,1.301) -- (-0.565,1.360) -- (-0.356,1.341) -- (-0.220,1.244) -- (-0.130,1.101) -- (-0.071,0.936) -- (-0.030,0.750) -- (-0.001,0.554) -- (0.025,0.354) -- (0.069,0.176) -- (0.188,0.064) -- (0.389,0.020) -- (0.600,-0.007) -- (0.810,-0.041) -- (1.001,-0.091) -- (1.162,-0.162) -- (1.290,-0.268) -- (1.361,-0.425) -- (1.347,-0.625) -- (1.280,-0.787) -- (1.189,-0.928) -- (1.082,-1.052) -- (0.962,-1.162) -- (0.825,-1.258) -- (0.663,-1.335) -- (0.453,-1.364) -- cycle;   
  \draw[gray!44] (0.374,-1.480) -- (0.191,-1.408) -- (0.070,-1.279) -- (-0.017,-1.117) -- (-0.084,-0.937) -- (-0.144,-0.749) -- (-0.213,-0.567) -- (-0.303,-0.412) -- (-0.433,-0.287) -- (-0.600,-0.199) -- (-0.790,-0.131) -- (-0.981,-0.069) -- (-1.167,0.006) -- (-1.324,0.103) -- (-1.437,0.238) -- (-1.482,0.443) -- (-1.441,0.643) -- (-1.359,0.812) -- (-1.260,0.959) -- (-1.143,1.094) -- (-1.012,1.217) -- (-0.869,1.324) -- (-0.702,1.416) -- (-0.507,1.477) -- (-0.272,1.453) -- (-0.125,1.349) -- (-0.022,1.200) -- (0.052,1.030) -- (0.114,0.843) -- (0.175,0.661) -- (0.252,0.489) -- (0.360,0.347) -- (0.507,0.242) -- (0.693,0.164) -- (0.887,0.100) -- (1.078,0.033) -- (1.251,-0.051) -- (1.388,-0.165) -- (1.471,-0.327) -- (1.469,-0.545) -- (1.405,-0.727) -- (1.314,-0.884) -- (1.205,-1.026) -- (1.079,-1.157) -- (0.942,-1.272) -- (0.789,-1.373) -- (0.609,-1.452) -- (0.379,-1.480) -- cycle;   
  \draw[gray!38] (0.306,-1.631) -- (0.091,-1.560) -- (-0.049,-1.426) -- (-0.154,-1.256) -- (-0.241,-1.070) -- (-0.325,-0.881) -- (-0.422,-0.705) -- (-0.546,-0.550) -- (-0.697,-0.427) -- (-0.879,-0.327) -- (-1.076,-0.238) -- (-1.266,-0.149) -- (-1.440,-0.038) -- (-1.569,0.106) -- (-1.631,0.318) -- (-1.598,0.549) -- (-1.515,0.744) -- (-1.412,0.910) -- (-1.290,1.066) -- (-1.153,1.211) -- (-1.006,1.340) -- (-0.842,1.457) -- (-0.662,1.555) -- (-0.443,1.624) -- (-0.192,1.608) -- (-0.017,1.500) -- (0.104,1.346) -- (0.198,1.165) -- (0.281,0.977) -- (0.370,0.794) -- (0.479,0.625) -- (0.615,0.487) -- (0.786,0.374) -- (0.975,0.283) -- (1.172,0.195) -- (1.353,0.099) -- (1.510,-0.028) -- (1.610,-0.198) -- (1.624,-0.443) -- (1.559,-0.652) -- (1.466,-0.828) -- (1.352,-0.990) -- (1.226,-1.137) -- (1.083,-1.275) -- (0.927,-1.399) -- (0.756,-1.508) -- (0.560,-1.594) -- (0.311,-1.631) -- cycle;   
  \draw[gray!38] (0.218,-1.773) -- (-0.015,-1.697) -- (-0.174,-1.551) -- (-0.291,-1.373) -- (-0.394,-1.178) -- (-0.498,-0.986) -- (-0.619,-0.806) -- (-0.766,-0.652) -- (-0.944,-0.523) -- (-1.138,-0.415) -- (-1.341,-0.309) -- (-1.529,-0.191) -- (-1.681,-0.040) -- (-1.768,0.171) -- (-1.753,0.429) -- (-1.676,0.643) -- (-1.569,0.836) -- (-1.442,1.012) -- (-1.304,1.173) -- (-1.152,1.324) -- (-0.986,1.463) -- (-0.808,1.586) -- (-0.606,1.693) -- (-0.370,1.765) -- (-0.091,1.748) -- (0.100,1.631) -- (0.235,1.466) -- (0.342,1.279) -- (0.443,1.083) -- (0.556,0.893) -- (0.688,0.727) -- (0.851,0.585) -- (1.040,0.467) -- (1.239,0.363) -- (1.436,0.254) -- (1.610,0.122) -- (1.735,-0.056) -- (1.773,-0.300) -- (1.719,-0.541) -- (1.626,-0.741) -- (1.509,-0.923) -- (1.376,-1.092) -- (1.231,-1.248) -- (1.071,-1.395) -- (0.898,-1.527) -- (0.712,-1.641) -- (0.497,-1.734) -- (0.223,-1.774) -- cycle;   


  \draw[-{Stealth[scale=0.8]}, thick, blue!60]
      plot[smooth] coordinates {(-1.05,-0.30) (-1.00,0.10) (-0.90,0.45) (-0.72,0.66)};
  \foreach \p in {(-1.05,-0.30), (-1.00,0.10), (-0.90,0.45)}
      {\fill[blue!70] \p circle (1.6pt);}
  \node[font=\tiny, blue!60!black, anchor=north] at (-1.05,-0.36) {A};

  \draw[-{Stealth[scale=0.8]}, thick, red!60]
      plot[smooth] coordinates {(0.90,0.50) (0.98,0.12) (0.95,-0.27) (0.85,-0.62)};
  \foreach \p in {(0.90,0.50), (0.98,0.12), (0.95,-0.27)}
      {\fill[red!70] \p circle (1.6pt);}
  \node[font=\tiny, red!60!black, anchor=south west] at (0.94,0.54) {B};

  \draw[-{Stealth[scale=0.8]}, thick, orange!75]
      plot[smooth] coordinates {(-0.38,0.40) (-0.18,0.18) (0.02,-0.06) (0.24,-0.34)};
  \foreach \p in {(-0.38,0.40), (-0.18,0.18), (0.02,-0.06)}
      {\fill[orange!85] \p circle (1.6pt);}
  \node[font=\tiny, orange!80!black, anchor=south west] at (-0.40,0.44) {C};
\end{tikzpicture}
};
\node[steplabel, above=0.0cm of step4] {4. Potential Landscape NN};

\draw[arrow] (step1) -- (step2);
\draw[arrow] (step2) -- (step3);
\draw[arrow] (step3) -- (step4);

\end{tikzpicture}
\caption{\rev{Pipeline for inferring the dynamical stance landscape. (1) Stance targets determined via corpus-oriented stance target extraction. (2) An LLM classifier determines stance (favor/neutral/against) for each document--target pair. (3) A latent Gaussian-process factor analysis model maps the sparse, irregularly sampled stance labels onto smooth low-dimensional trajectories $z(t)$ in a shared latent stance space, jointly handling temporal smoothing, missingness, and dimensionality reduction. (4) A potential landscape neural network learns the time-varying potential $\phi(z,t)$.}}
\label{fig:pipeline}
\end{figure*}

We \rev{apply} the landscape model from \citet{howe2025learning} to social media data containing stance expressions. We therefore need to process our text data into low-dimensional trajectories that we can use as input in our potential landscape model. Our pipeline is detailed in Fig. \ref{fig:pipeline}.

{\color{revcolor}%
We use corpus-oriented stance target extraction, which involves de-duplication, clustering, and canonicalization \citep{steel2025corpus} to find multiple stance-targets per document in our corpus (Step 1 in Fig. \ref{fig:pipeline}). We additionally translate any extracted stance targets that are not in English, into English, to consolidate stance expressions into fewer targets. See App. \ref{sec:target-extraction} for extra details.%
}

We then fine-tune an open-source large language model (LLM) \rev{\textit{Qwen3.5-4B}} \footnote{Apache 2 license} \citep{qwen3technicalreport} as a stance detection model to find the post's stance on these stance targets (Step 2 in Fig. \ref{fig:pipeline}). We train this model on VAST \citep{allaway2020zero}, EZ-STANCE \citep{zhao2024ez}, P-Stance \citep{li2021p}, SemEval \citep{mohammad2016semeval}, the multi-turn datasets MT-CSD \cite{niu2024multimodal} and CTSDT \citep{li2023contextual}, and the non-English language datasets Catalonia stance dataset \citep{zotova2020multilingual}, and the French election dataset \citep{evrard2020french}. The fine-tuned model achieved \rev{0.740} macro F1 on all datasets combined (more results in \rev{App. \ref{sec:stance-detection}),} and we released the model on HuggingFace \rev{\footnote{https://huggingface.co/bendavidsteel/Qwen3.5-4B-stance-detection}.} This gives us a large corpus of documents with stance classifications on their respective stance targets, and timestamps for each of those documents.

Why collect so many stance targets, as opposed to a select few? First, this gives us data efficiency: we are trying to record as many stance expressions as we can given our data. Second, it gives us redundancy, as one underlying attitude can be expressed via stance expressions on multiple issues \citep{feldman2014understanding}. We later reduce this redundancy via dimensionality reduction. Third, it lets us better observe changing attitudes: it may be that the political figures are reticent to explicitly renege \rev{on} their previous exact stance on a particular issue, but are happy to express their changed attitude on a similar issue, which we can then observe \citep{feddersen2023changing,andreottola2021flip}. \rev{We use all stance-targets with at least 400 posts, which results in 11,736 stance targets included, corresponding to 71\% of post-target pairs retained.} 

{\color{revcolor}%
We then use a latent Gaussian-process (GP) factor analysis model (GPFA) \citep{yu2008gaussian} (Step 3 in Fig. \ref{fig:pipeline}), which uses the stance classifications on their many targets to infer smooth, time-varying, low-dimensional trajectories, in a similar manner to previous work in latent ideology prediction \citep{barbera2015birds}. For our given stance trajectories for each public figure $m = 1,\dots,M$, talking about targets $j=1,\dots,J$, with posts at time $t$:
\begin{equation}
  f_{mjt} \;=\; \mathbf{w}_j^{\top}\mathbf{z}_m(t) + b_j
\end{equation}
Where $f_{mjt}$ is the sole predictor of the stance of a post at time $t$, $z_m(t)$ is a public figure's low dimensional trajectory, $\mathbf{w}_j$ is a vector mapping the low-dimensional trajectories onto the high-dimensional stance-targets, and $b_j$ is a vector setting the average level of each stance target. %
}

{\color{revcolor}%
Stance detection as a task is traditionally and most easily done as a classification task. However, the attitudes that stance detection infers lie on a continuous spectrum. Stance classifications are ordinal in nature: a `neutral' post favours a target more than an `against' post, and a `favor' post favours a target more than a `neutral' post. We therefore use an ordinal likelihood to map the predictor $f$ to stance probabilities. %
}
{\color{revcolor}%
Given that different aspects of stance may vary at different time-scales (i.e. beliefs that hold constant over one's life vs topical trends), we split the kernel dimensions into two sets: a fast set, and a slow one, with the number of fast dimensions ($n_{fast}$) and total dimensions ($n_{total}$) hyperparameters to optimize. We use a Wiener process kernel for the fast dimensions, and compare a Wiener kernel with a constant kernel for the slow dimensions. In order to train the GPFA, we bin posts into discretized cells specific to a figure-trajectory/time-bin/stance-target, where each bin carries the number of posts for each classification label, and where the bin temporal resolution is a hyperparameter. We determine the hyperparameters we use for the GPFA jointly with the potential landscape model as part of the hyperparameter optimization process we describe later. We include full details in App. \ref{sec:lgpfa}%
}

\citet{howe2025learning}'s method learns a potential landscape function that is purely a function of position in the latent landscape. The advantage of this approach is that it ties the shape of the potential landscape directly back to its predictive power. This method was used in a cellular development context, but here we apply it to stance changes. This is Step \rev{4} in Fig. \ref{fig:pipeline}. This approach uses the Langevin process equation, such that the next state of trajectories is dependent only on the current state. We assume that the system of stance change is over-damped \textemdash that is, that there is no inertia in the system, based on previous work showing that people generally change their stance due to new information in the same way \citep{coppock2023persuasion} (Although this has important \rev{caveats;} see App. \ref{sec:inertia} for an extended justification and discussion).  We model changes in the potential landscape over time, and therefore adapt our potential landscape function from \citet{howe2025learning} to be a function of time as well as position, modifying the update equation to:
\begin{equation}
    x'_{t+1} = x_{t} + \nabla\phi(x_t, t) + \eta
\end{equation}
Where \rev{$x_t$ is the trajectory position at time $t$, $x'_{t+1}$ is the predicted position of the trajectory at time $t+1$, $\phi(x_t, t)$ is the potential landscape, the gradient of which, $\nabla\phi$, is the model's predicted next trajectory movement, and $\eta$ is the diffusion characterizing movement by the trajectory not described by its descent of the potential landscape.} $\phi(x_t, t)$ is parameterized as a multi-layer perceptron (MLP), which outputs the scalar height of the potential landscape at that position and \rev{time. \citet{howe2025learning} track movement in trajectory distributions due to the inability to directly observe continuous trajectories from cells, and therefore use a distribution-based loss function. We have direct trajectories via the posterior mean from the GPFA, and therefore are able to use a mean-squared error (MSE) loss function, which provides a more direct method of measuring model accuracy. We have variable quantities of data over the landscape, so we use Monte-Carlo dropout to estimate the uncertainty of the landscape model at different points in the landscape \cite{gal2016dropout}. We use 10 samples from the model to determine the distribution for each point.}

{\color{revcolor}%
We hold out the final timespan of the dataset to use as a test-set, and for our forecasting evaluation, use a five-fold rolling holdout design. From the rest of the prior section of the dataset, we use 70\% of our trajectories for training, 10\% for validation, and hold-out 20\% of the trajectories for the final test set. This produces six sub-divisions on which to evaluate our model: three groups of trajectories that are either trained on, validated on, or tested, where predictive accuracy can be evaluated on the trained-on time period, or after the trained-on time-period. We train both our GPFA and landscape model only on the in-timespan stance classifications, except for the post-cutoff learned low-dimensional trajectories, which we infer using the frozen parameters and post-cutoff stance classifications. We adapt the GPFA in order to ensure that its trajectories prior to the cutoff do not use any knowledge of data past the cutoff. We determine the best model from a given run via loss on the validation dataset. We use Bayesian hyperparameter optimization via WanDB \footnote{https://docs.wandb.ai/models/sweeps, proprietary} to find the hyperparameters which maximize a metric indicating the extent to which the potential landscape is out-performing simple baselines; see App. \ref{sec:landscape-model} for details.%
}

\section{Experiments}
\label{sec:experiment}

\paragraph{Data}
\label{sec:datasets}

We use the dataset described in \citet{pehlivan2025can}. This is a dataset of posts from prominent Canadian accounts from the $1^{st}$ January 2022 \rev{until the $30^{th}$ June 2026,} across X/Twitter, TikTok, Instagram, and Bluesky. We use posts from accounts that could be reasonably assumed to be individuals \textemdash i.e. politicians and influencers, not civil society organizations or news organizations. Total post \rev{counts,} user counts for each \rev{platform, and breakdowns for figure type and political party} are shown in \rev{Tab.} \ref{tab:dataset}. A key consideration for this dataset is that the data \rev{comes} from political elites, not mass publics. Accounts were selected using a methodology defined in \citet{pehlivan2025building}, which requires that the accounts either belong to registered politicians, or influencers with a minimum 5000 follower count, and have the majority of their content be of a political nature. We pre-process data from each platform \rev{differently; see App.} \ref{sec:prepro}. When a person has accounts on multiple platforms, we use all posts across them to find their changing stance on each stance-target. \rev{Later on,} we look at individual accounts to compare platform movement.

\begin{table}
\centering
\begin{tabular}{lrr}
\toprule
Platform & Num Posts & Num Handles \\
\midrule
Bluesky & 1,411,667 & 930 \\
Instagram & 1,581,044 & 2,757 \\
Tiktok & 404,442 & 738 \\
Twitter & 22,897,605 & 3,640 \\
\midrule
Figure type & Num Posts & Num People \\
\midrule
Politician & 2,489,376 & 2,000 \\
Influencer & 22,915,478 & 2,298 \\
Foreign & 889,904 & 106 \\
\midrule
Party & Num Posts & Num People \\
\midrule
Conservative & 598,118 & 613 \\
Liberal & 462,751 & 401 \\
NDP & 438,005 & 402 \\
PPC & 86,564 & 145 \\
Bloc Québécois & 49,010 & 59 \\
Green & 43,670 & 135 \\
Independent & 12,409 & 3 \\
\midrule
Total & 26,294,758 & 4,404 \\
\bottomrule
\end{tabular}
\caption{\rev{Dataset statistics by platform, figure type, and party.}}
\label{tab:dataset}
\end{table}

\paragraph{Method Application}

{\color{revcolor}%
We evaluated our stance detection model against a stratified sample of manually coded post - stance target pairs. One of the authors manually annotated 444 post stance-target pairs from 250 posts, by looking at the posts in their platform contexts. 84\% of stance targets were coded as relevant to the post; see Tab. \ref{tab:coding-agreement} for stance detection statistics. As extra validation, we coded the post stance-target pairs with Claude Haiku 4.5 \footnote{https://platform.claude.com/docs/en/models/haiku-4-5/overview}, and obtain an inter-annotator agreement (Cohen's Kappa) of 0.69 with Haiku. The fine-tuned model obtains an F1 score of 0.743 against Haiku's annotations. We see variation in accuracy between the labels, with an F1 of 0.84 and 0.63 for the `favor' and `neutral' classes respectively. See App. \ref{sec:stance-detection} for class specific scores, calibration, and more subgroup statistics. While general agreement is high, our model performed poorly on pairs that include a transcript, meaning TikTok inferences should be treated with more caution.%
}

\begin{table}[t]
\begin{tabular}{@{}llrrr@{}}
\toprule
 & Subgroup & $n$ & $\overline{F}_1$ & $\kappa$ \\
\midrule
All posts & coded relevant & 375 & 0.73 & 0.62 \\
\midrule
Platform & twitter & 112 & 0.76 & 0.64 \\
 & instagram & 107 & 0.79 & 0.72 \\
 & tiktok & 85 & 0.60 & 0.41 \\
 & bluesky & 71 & 0.75 & 0.63 \\
\midrule
Year & 2022 & 58 & 0.77 & 0.65 \\
 & 2023 & 84 & 0.70 & 0.57 \\
 & 2024 & 82 & 0.76 & 0.66 \\
 & 2025 & 80 & 0.65 & 0.57 \\
 & 2026 & 71 & 0.76 & 0.63 \\
\midrule
\raisebox{0pt}[0pt][0pt]{\begin{tabular}[t]{@{}l@{}}Political \\ group\end{tabular}} & Liberal & 36 & 0.65 & 0.46 \\
 & Conservative & 34 & 0.59 & 0.48 \\
 & NDP & 27 & 0.96 & 0.92 \\
 & Bloc Québécois & 38 & 0.66 & 0.49 \\
 & Green & 37 & 0.83 & 0.78 \\
 & Other party & 50 & 0.71 & 0.57 \\
 & Independent & 35 & 0.77 & 0.71 \\
\midrule
\raisebox{0pt}[0pt][0pt]{\begin{tabular}[t]{@{}l@{}}Actor \\ type\end{tabular}} & politician & 290 & 0.74 & 0.62 \\
 & influencer & 58 & 0.77 & 0.64 \\
 & foreign & 27 & 0.61 & 0.45 \\
\midrule
Language & English & 261 & 0.72 & 0.62 \\
 & French & 94 & 0.69 & 0.54 \\
\midrule
\raisebox{0pt}[0pt][0pt]{\begin{tabular}[t]{@{}l@{}}Text \\ provenance\end{tabular}} & written & 286 & 0.78 & 0.69 \\
 & transcript (ASR) & 59 & 0.51 & 0.28 \\
 & image text (OCR) & 30 & 0.70 & 0.55 \\
\bottomrule
\end{tabular}
\caption{\rev{Comparing the post stance-target pairs classified by our stance detection model with a manually annotated set of pairs, reporting agreement via macro F1 and Cohen's Kappa. Note that the number of English and French posts does not sum to the total number of relevant stance-targets, as there were posts in other languages. We fold posts from PPC politicians into `Other party' to save space in this table; see more details in the appendix.}}
\label{tab:coding-agreement}
\end{table}

{\color{revcolor}%
We used July 2025 onwards as a cut-off point for our hold-out test-set. As a measure of the goodness-of-fit of the final best GPFA used in the rest of the analysis, we look at the percentage of achievable discrimination on stance direction that it obtains. This is 52\% for in-train-time trajectory sections, and 44\% in the held-out temporal portion. We tested the stability of the low-dimensional space the GPFA learns by repeated re-training on disjoint sets of people, and found that the first three dimensions were stable (a greater than 0.9 cosine score on similarity of the principal angles between subspaces) across these re-trainings. We therefore focus on the first three dimensions in the rest of this analysis. We include more details, and a figure of the raw stance trajectories, in App. \ref{sec:lgpfa}. Without the interpretive layers we add later, they are an uninterpretable mess of lines.%
}

{\color{revcolor}%
The hyperparameter optimization process led to a homogeneous kernel of all fast kernel priors (that is, $n_{fast} = n_{total}$ for our final model). We describe the final parameters used as determined by hyperparameter optimization sweeps in App. \ref{sec:landscape-model}. \footnote{We complete experiments on two NVIDIA RTX 3090s and a research cluster of H100s.}%
}

\paragraph{Stationarity}

{\color{revcolor}%
We test for the stationarity of our system with a variety of methods, using a combination of model-free tests, and tests on the trajectories produced by the GPFA, all on a 16-day grid. For a model-free test, we use a variogram, which gives a drift standard deviation of 0.11, indicating drift in the data. The rise in excess variance continues at the longest lag times, indicating the drift is unbounded, and not saturating as bounded, mean-reverting motion would.%
}

{\color{revcolor}%
We then run tests on the time-varying trajectories produced by the GPFA, balanced to ensure the test is not distorted by unbalanced time-spans of people's posting patterns, leaving a panel of 1005 figures over 99 bins. We compute a rolling-window mean and variance over the data \citep{hyndman2018forecasting}, dividing the system into 6 windows. The maximum mean drift of the dimensions is +0.195 as measured using Cohen's d, which is considered a small effect size according to Cohen's d guidelines \citep{cohen2013statistical}. The maximum variance drift relative to the initial variance is -18\%. These results are inconsistent with weak stationarity. We additionally do a Zivot-Andrews test \cite{zivot2002further}, and find breaks in September 2024 and January 2025 in the first and second dimensions respectively. We also test whether the ensemble disperses, measuring each trajectory's distance from the precision-weighted centroid at each time point. The mean distance changes by a factor of 1.03 across the time period. We report additional statistical tests in App. \ref{sec:app-station}.%
}

Taken together, these results indicate that the system is drifting, not \rev{dispersing, and that} the drift is \rev{punctuated by structural breaks rather than stationary around a fixed landscape. We cannot say for certain whether the data is better modelled by mean-reversion on a timescale longer than our observation window, or} a random walk. This indicates that modeling the system with a potential landscape that does not assume stationarity is the right choice. However, future work should verify this with more data, and explore alternative modelling choices.

\paragraph{Landscape Model Fit}
\label{sec:model-fit}

\begin{figure}[t]
    \centering
    \includegraphics[width=\linewidth]{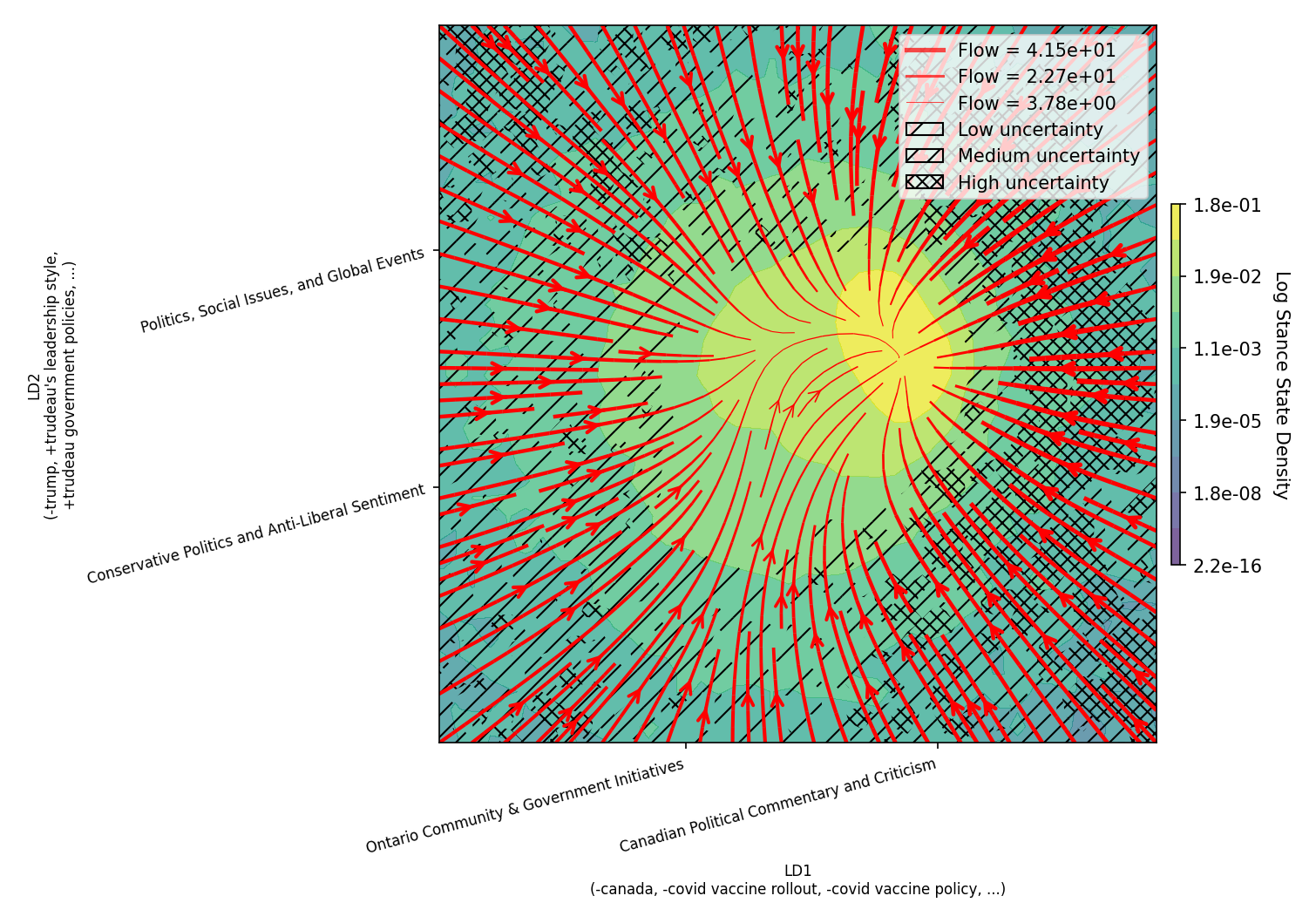}
    \caption{\rev{Gaussian KDE of trajectory density with streamplots from the potential landscape model. Hatching shows model uncertainty, shown on the first two dimensions of the latent stance space. Lighter coloured shades of the colormap show the density of trajectory points at that point in the landscape, and streamplot line direction and width show the direction and strength with which we infer that trajectories in that region of the landscape move. Therefore from this figure, we can infer that, in general, people's trajectories in stance space converge on the modal stance point, and that there is an attractor like in Figure \ref{fig:explanatory}, but not a repellor. We show the top stance-targets loaded on by each dimension as axis labels, and topic descriptors of the language used at specific points in the space as tick labels.}}
    \label{fig:density-and-streams}
\end{figure}

We show the same first two dimensions of the space, with kernel density estimation and the streamplot showing the gradient of the potential landscape as descriptive layers on top of the trajectories, in Fig. \ref{fig:density-and-streams}. For future visualizations, we show only the streamplots from the landscape model for increased clarity.

\begin{figure*}[t]
    \centering
    \includegraphics[width=0.8\linewidth]{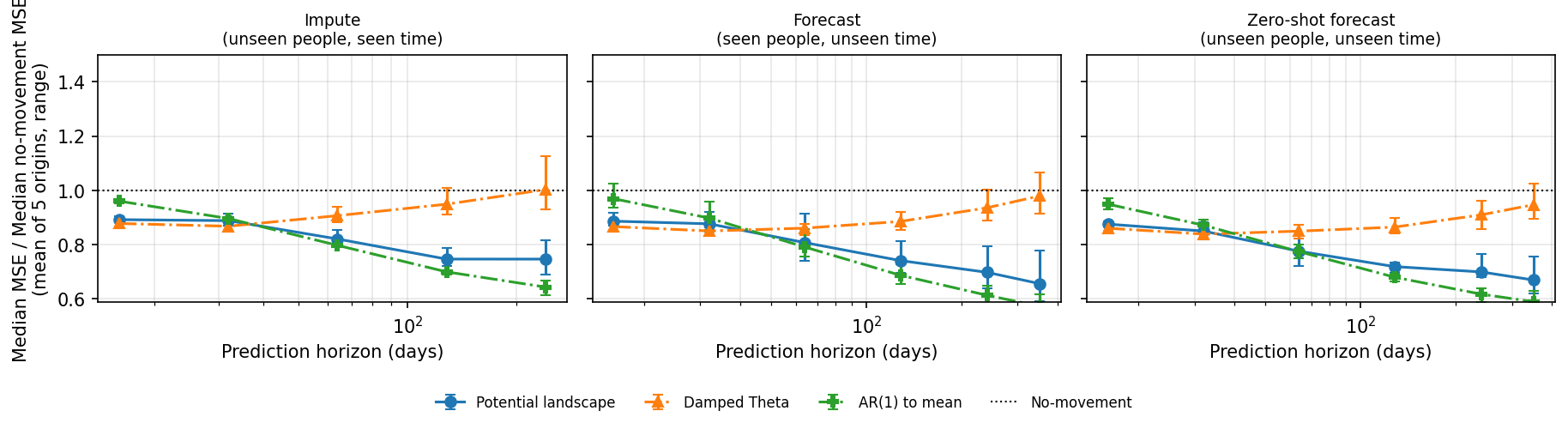}
    \caption{\rev{Potential landscape predictive-ness on held-out trajectories as compared to the predictive power of the baseline that trajectories stay stationary, alongside two comparison time-series forecasting methods. Normalized by median MSE of baseline at the same horizon. Lower is better. Our model out-performs the baseline on all time horizons.}}
    \label{fig:predictive-horizon}
\end{figure*}

{\color{revcolor}%
We use a five-fold rolling holdout to report the in-train-time test-set scores, the out-of-train-time train-set scores, and out-of-train-time test-set scores, which demonstrate the interpolative, forecasting, and zero-shot forecasting capability of the landscape model respectively. By comparing our model to a naive forecast of `trajectories don't move in stance space', our model obtains a median MSE improvement over the baseline model of 33\%, at 360 days out, when doing zero-shot forecasting (untrained on figures in the heldout temporal test period); see Fig. \ref{fig:predictive-horizon}. We can see that the potential landscape is better than the baseline at all time-scales we measure. We compare this to damped-Theta and auto-regressive mean reversion time-series prediction models on the same trajectories \citep{assimakopoulos2000theta}, again compared against the simple no movement baseline. %
}

{\color{revcolor}%
The landscape model is more predictive than the damped-Theta model at a horizon of more than 30 days, and is more predictive than the mean reversion model below 60 days. We infer that the landscape model is out-performed by the auto-regressive mean reversion model because, over long horizons, the dominant pattern of the data is global shrinkage towards the centroid. However, our landscape model localizes where and when that flow is stronger, which lets it out-perform that method on short horizons.%
}

{\color{revcolor}%
The model is significantly ($p < 0.05$ paired Wilcoxon signed-rank) better at predicting trajectories of all figure types than the baseline. It is significantly (by the same test) better at predicting Liberal, NDP, Conservative, and PPC politician trajectories than the baseline, and we don't have enough data for Bloc Québécois or Green party politicians in our test set to reliably say either way.%
}

\begin{figure}
    \centering
    \begin{subfigure}{0.8\linewidth}
        \centering
        \includegraphics[width=0.9\linewidth]{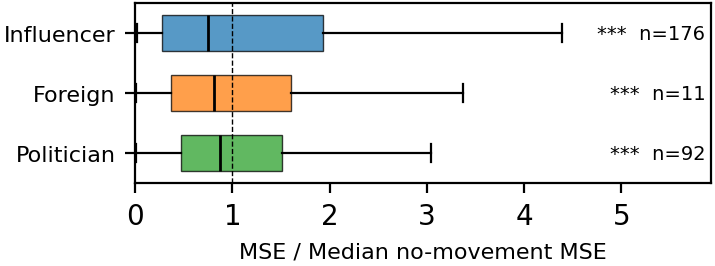}
        \caption{\rev{By figure type.}}
        \label{fig:maintype-perf}
    \end{subfigure}
    \begin{subfigure}{\linewidth}
        \centering
        \includegraphics[width=0.8\linewidth]{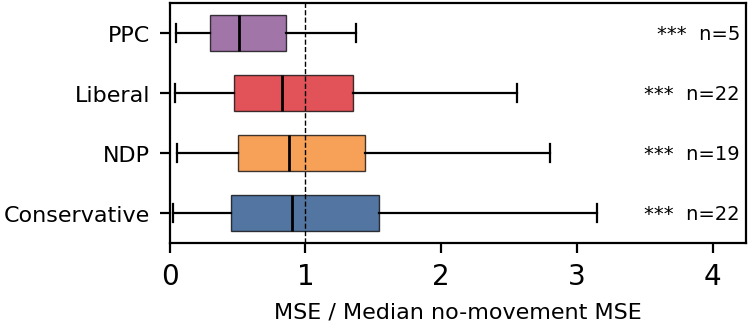}
        \caption{\rev{By party (politicians only). We remove parties for whom we have less than five trajectories in our test set.}}
        \label{fig:party-perf}
    \end{subfigure}
    \caption{\rev{Predictive performance breakdowns at the 30-day horizon. (a) The model is significantly better than the baseline at predicting all figure types, and appears to be better at predicting influencer trajectories than politician trajectories. (b) The model is significantly better than the baseline at predicting the trajectories of politicians from parties for whom we have at least five trajectories in the test set.}}
    \label{fig:eval-breakdowns}
\end{figure}

\paragraph{Dimensions}
\label{sec:dims}

We show a representation of the \rev{latent dimensions (LDs) discovered by the GPFA in our dataset, in Tab.} \ref{tab:dims}. To provide descriptions of the text in each percentile range, we use Toponymy (a topic-modelling library) \footnote{https://github.com/TutteInstitute/toponymy, MIT license} to identify keyphrases associated with the texts, pick random exemplars, and then use Gemma 4 E4B instruct \footnote{https://huggingface.co/google/gemma-4-E4B-it, Apache 2 license} to summarize the texts. 

\begin{table*}[t]
    \centering
    \scriptsize
    \begin{tabularx}{\textwidth}{c|c|XX|XXXX}
    \toprule
    & & & & \multicolumn{4}{c}{\textbf{Description}} \\
    \cmidrule(lr){5-8}
    \textbf{Dim.} & \textbf{Var.} & \textbf{Targets ($+$)} & \textbf{Targets ($-$)} & \textbf{0--1\%} & \textbf{1\% -- 10\%} & \textbf{90\% -- 99\%} & \textbf{99\% -- 100\%} \\
    \midrule
           1 & 49\% & trump,\newline russian foreign policy,\newline russian government influence,\newline russian military actions & canada,\newline covid vaccine rollout,\newline covid vaccine policy,\newline indigenous immigration policies & Government Initiatives and Community Growth & Local Governance and Economic Growth & Culture Wars, Politics, and Extremism & Conservative Anti-Establishment Commentary \\
           2 & 18\% & trudeau's leadership style,\newline trudeau government policies,\newline trudeau's cabinet decisions,\newline joe biden & trump,\newline donald trump,\newline israel,\newline pierre poilievre & Canadian Conservative Activism \& Populism & Cross-Border Conservative Policy Analysis & Politics, Climate, and COVID-19 Issues & Health, Climate, and Social Justice \\
           3 & 13\% & covid vaccine rollout,\newline covid vaccine access,\newline women,\newline covid vaccine policy & trudeau's leadership style,\newline trudeau government policies,\newline trudeau's cabinet decisions,\newline joe biden & Federal \& Provincial Governance Issues & Online Political Commentary \& Culture & Health and Political Activism & National \& Provincial Policy Debates \\
        \bottomrule
    \end{tabularx}
    \caption{\rev{Description of the first three latent dimensions (LDs) extracted from the stance observations by the GPFA. The stance targets column shows the top four most important stance targets for each loading, for both positive ($+$) and negative ($-$) loadings. The description columns show a summary description of the posts that result in a user being at a specific percentile of that dimension. We obtain these descriptions by finding trajectory sections that were in these specific percentile ranges at any point in time, finding the posts they made that are linked to stance-targets that are in the top 30 loadings of that specific LD, and passing these posts to the topic naming functions available in Toponymy (see main text for footnote). We also show the percentage of the total variance in stance-target space captured by the model that each LD explains.}}
    \label{tab:dims}
\end{table*}

{\color{revcolor}%
To understand the differences between the first three LDs, we computed Cohen's d between pairs of groups based on our dataset metadata, for the first three LDs. See Tab. \ref{tab:dim-group-cohens}. %
}

\begin{table}[t]
    \centering
    \begin{tabular}{l|ccc}
    \toprule
    \textbf{Contrast} & \textbf{LD1} & \textbf{LD2} & \textbf{LD3} \\
    \midrule
        Influencer vs.\ politician & $\mathbf{+2.2}$ & $+0.6$ & $0.0$ \\
        CPC.\,$+$\,PPC vs.\ NDP\,$+$\,Green & $+0.4$ & $\mathbf{-2.8}$ & $0.0$ \\
        LPC vs.\ all other MPs & $-2.0$ & $+1.6$ & $\mathbf{-3.4}$ \\
        Gov.\ vs.\ opposition MLAs & $-1.3$ & $-1.1$ & $\mathbf{-1.5}$ \\
        PPC vs.\ all other parties & $\mathbf{+1.9}$ & $-0.8$ & $+1.2$ \\
        \bottomrule
    \end{tabular}
    \caption{\rev{Cohen's d between pairs of groups, for each dimension. Highest absolute value on each row is bolded.}}
    \label{tab:dim-group-cohens}
\end{table}

{\color{revcolor}%
From the results in this table, combined with Tab. \ref{tab:dims} above, we infer that the first LD is a generalized populist/institutional dimension, given that it is the dimension where the politician centroid and the influencer centroid are most split. In addition, the politicians of the People's Party of Canada (PPC), a right-wing populist party, are most separated from the politicians of the other, non-populist parties on this dimension. We infer the second dimension to be a Canadian left-right party dimension, given that this is the dimension where the centroid of Conservative Party of Canada (CPC) and PPC politicians (the two largest right-wing parties of Canada), and the centroid of the New Democratic Party (NDP) and Green party politicians (the two largest left-wing parties of Canada), are the most separated. And finally, we infer that the third dimension is a Canada-specific government/opposition dimension, given that this is the dimension with the largest distance between the Liberal Party of Canada (LPC) politicians (the governing party of Canada since 2015), and all other parties, and the dimension with the largest distance between governing provincial members of the legislative assemblies (MLAs, provincial politicians), and opposition MLAs. We include more detail in App. \ref{sec:dim-analysis}.%
}

\citet{lupton2015political} indicate that \rev{politicians'} attitudes are more uni-dimensional than others. When we look at the \rev{eigenvalues of the between-user covariance for} politicians vs influencers (essentially a measure of the dimension concentration of the two groups), \rev{there is no significant difference in the $\lambda_1$ between the groups:  0.314 and 0.304 for} politicians and influencers respectively. This may be explained by \citet{lupton2015political}'s US-focused dataset, \rev{drawn from} a more polarized society with only two dominant political parties, compared to Canada's larger set of political parties.

\paragraph{Time}

\begin{figure*}[t]
    \centering
    \includegraphics[width=\linewidth]{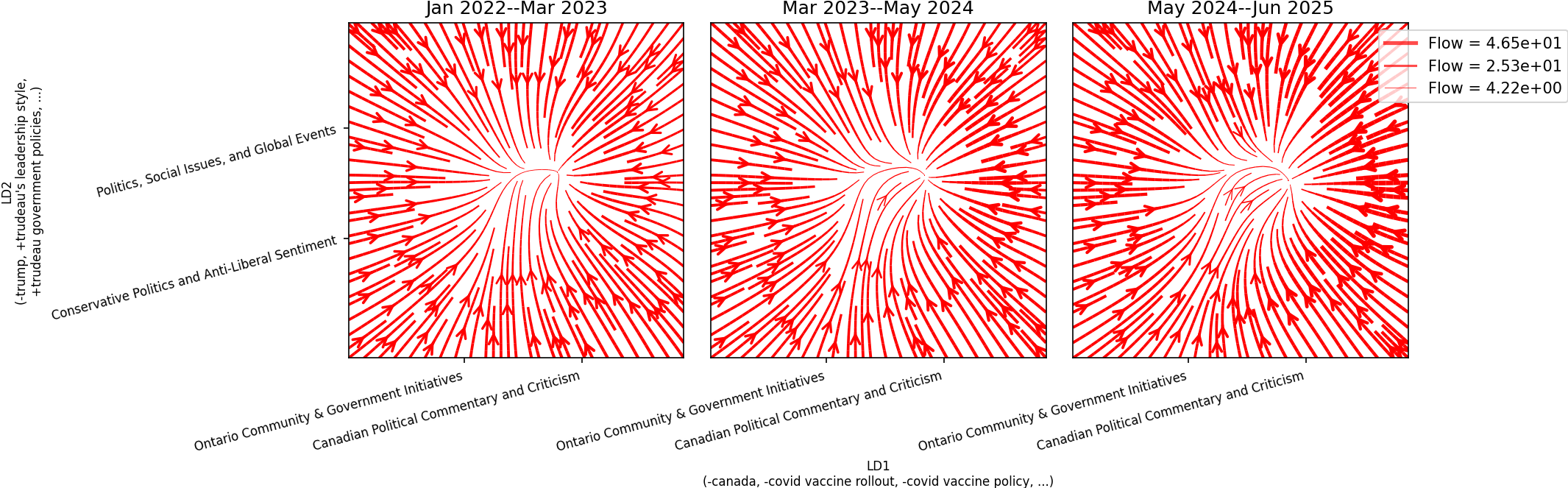}
    \caption{\rev{Snapshots in time of the evolving potential landscape, for the first two latent dimensions. The attractor point in the figure moves nine percentile points higher on the first LD, and 10 percentile points lower on the second LD, in addition to a 43\% increase in max flow over the periods shown.}}
    \label{fig:snapshots}
\end{figure*}

Using time as a parameter of our model allows us to see how the potential landscape moves over time, and we show this evolution through the years in Figure \ref{fig:snapshots}.

{\color{revcolor}%
We see one attractor point that moves over the three periods shown in the figure, moving from the 63rd to the 72nd percentile on the first LD, and the 55th to the 45th percentile on the second LD over the period. Going back to our dimensional analysis in Sec. \ref{sec:dims}, this implies the attractor is moving more populist, and more right-wing. Note however that due to the high proportion of Twitter data in the dataset, this may be mostly due to Twitter dynamics; see the Platforms section. The movement on the second LD is equivalent to 24\% of a figure's own median one-year movement. There is a 43\% increase in max flow between the periods. %
}

{\color{revcolor}%
We see general centralizing flows in both of the first two latent dimensions, and the movement of the attractor higher on the first LD, and lower on the second LD. What does this mean in practice? To put the movement that the potential landscape captures into perspective, we look at the people who move most in either direction for each dimension, and find the specific loaded stance-targets that they make significant moves on, as an exploratory measure (with BH correction; see App. \ref{sec:stance-examples} for details on statistical tests). In 2022, we see the top ten percent of positive movers on the first LD post less content favoring, and more content against, `canada'. In 2023, the corresponding most moved on stance target for these positive movers on the first LD is `indigenous immigration policies', in 2024, it is `kamala harris', and in 2025 it is `canada' again. For the second LD, in 2022, the top ten percent of negative movers on the second LD posted more content in favor, and less against, `israel', and `trump'. In 2023, they posted less content in favor of, and more content against, `indigenous immigration policies' and `vaccine'. In 2024, the top negative movers posted less content in favor of `joe biden', and the equivalent stance target in 2025 is `palestinian'. We include tables showing more significant examples in App. \ref{sec:stance-examples}.%
}

\paragraph{Platforms}
\label{sec:platforms}

\begin{figure*}[t]
    \centering
    \includegraphics[width=\linewidth]{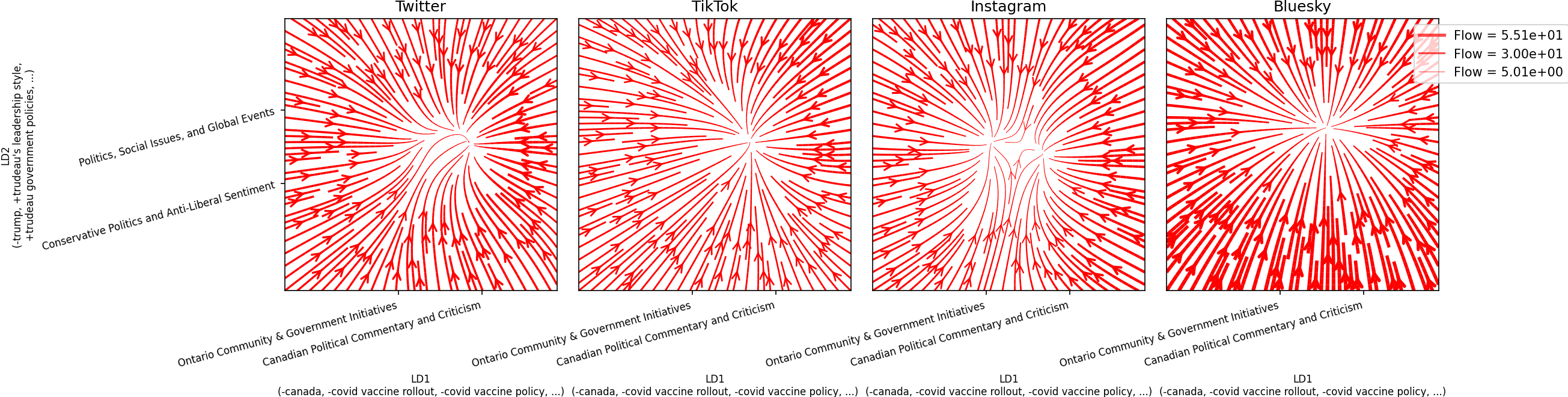}
    \caption{Comparison of the potential landscape of the 4 platforms in the dataset. We do dimensionality reduction on all account trajectories together, then train individual potential landscape models on the user trajectories of each platform. Note that the difference in streamplot strength between the platforms may not be due to genuinely stronger trajectory movement, but instead due to different relative quantities of data on each platform, resulting in individual trajectories affecting the landscape more.}
    \label{fig:platform-landscapes}
\end{figure*}

Given that we have data from multiple platforms in our dataset, we can use this to unpack the contributions of each platform to the potential landscape we found when looking at the platforms together. To separate platform potential landscapes, we fit \rev{the model only on posts from each specific platform.} We use the saved \rev{target loadings} model from the default pipeline so that the dimensions use the same \rev{loadings} as for the default pipeline (see additional details in App. \ref{sec:landscape-model}). \rev{We see that the potential landscape is most predictive on TikTok (27\% improvement over baseline, although this is likely due to less data from TikTok on our dataset), and least predictive on Bluesky (11\% improvement over baseline).} See the results in Fig. \ref{fig:platform-landscapes}.

{\color{revcolor}%
We see that the attractor point is in slightly different places across the four platforms, and see two attractor points for Instagram, with a saddle point between them. By describing attractor position as a percentile of the pooled cloud of all trajectory points, we can see that on the first LD, the range stretches from Instagram's lower attractor (on LD1) at the 19th percentile to X/Twitter's attractor at the 74th percentile. On the second LD, the range stretches from Instagram's other attractor at the 35th percentile to Bluesky at the 70th percentile. Referring back to our dimension interpretations in Sec. \ref{sec:dims}, this implies Twitter is more populist/anti-institutional, whereas one of Instagram's attractors is more institutional. And on the second LD, Bluesky is more left-wing, passing a sanity check.%
}

{\color{revcolor}%
We see general convergent movement on both LDs on all platforms. For Twitter, we see more against stances on `transgender rights' and `covid vaccine policy' among people moving in the positive direction of the first LD, but among negative movers on that dimension we see more content favoring `kamala harris'. On the second LD among positive movers, we see more content against `israel' and `elon musk'. On TikTok, among positive movers on the first LD, we see more content against `mark carney' and `climate change policy', but among negative movers less content favoring and more content against `trump'. On Instagram, among positive movers on the first LD, we see more content against, and less content for, `abortion'. And among negative movers on the second LD, we see more content against `liberal party policies'. On Bluesky, among positive movers, we see less content favoring, and more content neutral, on `indigenous rights', whereas among negative movers on that LD, we see more content favoring `covid vaccine policy'. We give these examples to indicate the kind of specific stance changes that contribute to the large-scale dynamics we see in the figure.%
}

{\color{revcolor}%
We include a comparison of these results to a survey on the Canadian general public over the same years; see App. \ref{sec:survey-results}.%
}

Note that the difference between platforms we show here does not let us comment on the causal effect of participating in the platform. We are not controlling for the self-selection of users of each platform. However, it does let us comment on the output content of the combination of platform and user-base, which is valuable in its own right.

\section{Discussion}
\label{sec:discussion}

\paragraph{Limitations}
\label{sec:limitations}

{\color{revcolor}%
The way that people present their stances changes over time. Classifiers that work in one year may perform worse in other years \cite{lazaridou2021mind}. Therefore models that rely on supervised fine-tuning may not be the ideal approach. However, based on our stratified sample of manual annotations across the years in the dataset (App. \ref{sec:stance-detection}), there is no monotone trend in stance detection F1, so this is at least not evident in our case.%
}

{\color{revcolor}%
We use noun phrases as stance targets, which can be unclear (i.e. what does it mean to `favor' `unemployment rate'?). We trialled using claims as stance targets, which do produce more specific targets, but this comes at the cost of much fewer posts per stance target (1925 stance targets extracted are mentioned over a thousand times for noun phrases, against only 24 stance targets for claims), therefore increasing the computational resource requirements of the modelling work due to the sparser stance targets. We therefore used noun-phrases in this work, but future work could revisit this choice. We also do not use direct visual or audio data to determine stance, instead using optical character recognition and audio transcripts. The lack of use of this data leads us to more poorly characterize stance on platforms where the content is multi-modal i.e. TikTok and Instagram.%
}

There are many assumptions we have made about the dynamical landscape here (i.e. the Markov assumption), and we must refine \rev{them} as this work progresses.

\paragraph{Future Work}
\label{sec:future-work}

Our method provides insight into the nature of large stance movements. While previous work has shown that partisans set the agenda for politicians in terms of the topics they talk about \citep{barbera2019leads}, this method can show who changes their expressed views first and last between partisans and politicians. \rev{Future work could also more explicitly model discrete events or information, to understand how they change the stance potential landscape.}

We \rev{could} experiment with non-linear dimensionality reduction techniques, to see if it leads to higher predictive accuracy. However, using non-linear techniques would require adaptations to the potential landscape model due to use of non-Euclidean distances. Future work could also \rev{relax} the Markov assumption, in order to try to increase predictive-ness.

\section{Conclusion}

{\color{revcolor}%
We introduce a method to provide a low-dimensional descriptive map of the mass change in publicly expressed stance expressions of Canadian political elites. We show that the low-dimensional space in which the movement occurs is highly explainable, and suggests a rich new method of exploring the dimensions of political stance. We confirm that the inferred potential landscape used to describe it is also comparable in predictive power to time-series forecasting methods, while showing detailed local dynamics. In particular, those dynamics reveal attractors in stance space that move slowly over time, and vary by platform, in a manner that passes sanity checks based on common knowledge of the platforms. We note that these results showing attractors, and not repellors, run in opposition to much recent literature on polarization, and hope that this indicates that this method can provide a rich new way of describing platform dynamics.%
}

\bibliography{sources}

\appendix

\section{Ethical Statement}
\label{sec:ethics}

In this work, we specifically focus on creating descriptive models of general behaviour, rather than focusing on focused models of individual behaviour, in order to avoid the potential mis-use of such a method to be used for opinion manipulation.

Additionally, we also focus on prominent public figures as opposed to members of the general population, because there is already an expectation of public engagement with their produced content, and therefore a reduced degree of privacy violation.

The development and use of stance detection as part of this work however could broaden the knowledge of this method to potential bad actors looking to understand stances so that they can be influenced. However, we believe this ability is already widely known (and widely implemented in a stronger form in social platforms), and therefore we believe this risk to be extremely marginal.

\section{Text Preprocessing}
\label{sec:prepro}

We needed to pre-process all social media posts into a common format so that our stance detection model can uniformly process them.

We use whisper to transcribe the audio from our TikTok videos. In this process we also obtain speaker embeddings. We cluster these embeddings using DBSCAN \citep{ester1996density} to find speaker clusters across all TikTok videos. We set the threshold of the DBSCAN clustering based on a small dataset of annotated speaker pairs. Based on these speaker clusters, if a speaker features in at least 40\% of an account's posts, then we consider them an author of that account. We then format the transcripts such that authors are unquoted, but all other speakers are quoted. This is to make TikTok text content similar to text-based platform content, where posters may quote other people.

{\color{revcolor}%
For Instagram, we extract the text of the images via optical character recognition (using EasyOCR \footnote{https://github.com/jaidedai/easyocr}), and use high-confidence text extractions ($\ge 0.75$), in addition to the caption of the post as the text content. And for X/Twitter and Bluesky, we use the textual content of the post, plus previous post content that the post is replying to.%
}

\section{Stance Target Extraction}
\label{sec:target-extraction}

{\color{revcolor}%
We use an embedding de-duplication distance of 0.15, on embeddings from \textit{intfloat/multilingual-e5-small} \footnote{https://huggingface.co/intfloat/multilingual-e5-small}. We use HDBSCAN clustering with a minimum cluster size of 200, min samples of 10, and the \textit{leaf} cluster selection method. We use the \textit{lingua} library \footnote{https://github.com/pemistahl/lingua-py} to determine the language of targets, and use a \textit{Qwen3.5-4B} model (reasoning disabled) to translate the targets into English.%
}

{\color{revcolor}%
We use the following prompt:
\begin{quote}\small
The phrase below is a topic label taken from a social media post. If it is not in
English, translate it into English. If it is already English, reply with it
unchanged. Keep proper names, place names, organisations, publications and
abbreviations as they are normally written in English instead of translating them
word by word. Do not begin your reply with ``the'', ``a'' or ``an''. Reply with
only the phrase: no quotes, no explanation, no notes.

\medskip
\noindent Phrase: \texttt{\{target\}}
\end{quote}%
}

\section{Stance Detection}
\label{sec:stance-detection}

Fine-tuned stance detection evaluation results are shown in \rev{Tab.} \ref{tab:stance-finetune}. Note that P-Stance does not include `neutral' labels, only `favor' and `against' labels. All other datasets contain all labels. We still include P-Stance purely for a larger diversity of \rev{training} data.

\begin{table}[]
    \centering
    \begin{tabular}{c|ccc}
    \toprule
        Dataset & F1 & Recall & Precision \\
    \midrule
        VAST & 0.78 & 0.78 & 0.78 \\
        EZ-STANCE & 0.67 & 0.66 & 0.68 \\
        P-Stance & 0.85 & 0.85 & 0.85 \\
        SemEval & 0.69 & 0.72 & 0.70 \\
        MT-CSD & 0.63 & 0.61 & 0.66 \\
        CT-SDT & 0.75 & 0.72 & 0.79 \\
        Catalonia & 0.75 & 0.75 & 0.75 \\
        French & 0.70 & 0.82 & 0.68 \\
    \bottomrule
    \end{tabular}
    \caption{Evaluation results of the fine-tuned stance detection model we used in this work.}
    \label{tab:stance-finetune}
\end{table}

{\color{revcolor}%
We manually annotated 250 posts from a stratified sample of the pairs of posts and stance targets, corresponding to 444 post/stance-target pairs. We report the target extraction agreement statistics in Table \ref{tab:target-extraction}. We report the statistics as compared to the stance detection model in Table \ref{tab:stance-agreement}, the comparison with Claude Haiku 4.5 annotations in Table \ref{tab:stance-claude}, and additionally the label specific breakdowns of precision and recall in Table \ref{tab:stance-class}. We show that the model is well calibrated in its prediction probabilities in Table \ref{tab:stance-confidence}. We report on the subgroup classifier error in Table \ref{tab:subgroup-error-rates}.%
}

{\color{revcolor}%
We do not report on skewed/biased stance targets, as we are doing zero-shot stance detection, and therefore the set of stance-targets used in the fine-tuning dataset is not the same as the set of stance-targets focused on in the final analysis.%
}

\begin{table*}[]
\centering
\begin{tabular}{lrrr}
\toprule
 & $n$ & \% & 95\% CI \\
\midrule
Extracted targets judged relevant & 444 & 84.5 & [80.9, 87.8] \\
\quad weighted to all extracted pairs & 444 & 78.1 & [72.6, 83.0] \\
Posts with at least one relevant target & 250 & 92.8 & [88.9, 95.4] \\
Mean per-post share of relevant targets & 250 & 85.0 &  \\
\bottomrule
\end{tabular}
\caption{\rev{Statistics showing the rate at which targets matched to posts are relevant to that post.}}
\label{tab:target-extraction}
\end{table*}

\begin{table*}[]
\centering
\begin{tabular}{lrr}
\toprule
 & All coded pairs & Coded relevant \\
\midrule
$n$ (pairs) & 444 & 375 \\
Accuracy / micro-F1 & 0.748 [0.707, 0.791] & 0.763 [0.717, 0.805] \\
Macro-F1 & 0.736 [0.693, 0.780] & 0.734 [0.686, 0.780] \\
Weighted F1 & 0.744 [0.704, 0.788] & 0.759 [0.714, 0.802] \\
Cohen's $\kappa$ & 0.606 [0.544, 0.671] & 0.617 [0.548, 0.682] \\
\bottomrule
\end{tabular}
\caption{\rev{Measures of the extent to which our annotator agree with our fine-tuned stance detection model on a stratified sample of classifications.}}
\label{tab:stance-agreement}
\end{table*}

\begin{table*}[]
\centering
\begin{tabular}{@{}lrrrr@{}}
\toprule
 & $n$ & Acc. & $\overline{F}_1$ & Cohen's $\kappa$ \\
\midrule
Classifier vs.\ Author & 444 & 0.748 & 0.736 & 0.606 [0.544, 0.671] \\
Classifier vs.\ haiku-4-5 & 444 & 0.748 & 0.743 & 0.609 [0.544, 0.673] \\
\midrule
Author vs.\ haiku-4-5 & 444 & 0.804 & 0.808 & 0.692 [0.633, 0.754] \\
\bottomrule
\end{tabular}
\caption{\rev{Comparison of our annotations with Claude Haiku 4.5.}}
\label{tab:stance-claude}
\end{table*}

\begin{table*}[]
\centering
\begin{tabular}{lrrrrrrrr}
\toprule
 & \multicolumn{4}{c}{All coded pairs} & \multicolumn{4}{c}{Coded relevant} \\
\cmidrule(lr){2-5} \cmidrule(lr){6-9}
Class & P & R & F1 & $n$ & P & R & F1 & $n$ \\
\midrule
Favor & 0.761 & 0.857 & 0.806 & 189 & 0.827 & 0.857 & 0.842 & 189 \\
Against & 0.647 & 0.786 & 0.710 & 84 & 0.688 & 0.786 & 0.733 & 84 \\
Neutral & 0.806 & 0.608 & 0.693 & 171 & 0.699 & 0.569 & 0.627 & 102 \\
\midrule
Macro & 0.738 & 0.750 & 0.736 & 444 & 0.738 & 0.737 & 0.734 & 375 \\
\bottomrule
\end{tabular}
\caption{\rev{Per class stance detection precision, recall, and F1, for all coded pairs, and only for those coded relevant.}}
\label{tab:stance-class}
\end{table*}

\begin{table*}[]
\centering
\begin{tabular}{lrrrr}
\toprule
Confidence & $n$ & Mean conf. & Accuracy & Gap \\
\midrule
0.3--0.4 & 1 & 0.373 & 0.000 & -0.373 \\
0.4--0.5 & 24 & 0.465 & 0.250 & -0.215 \\
0.5--0.6 & 45 & 0.555 & 0.556 & +0.001 \\
0.6--0.7 & 48 & 0.653 & 0.604 & -0.049 \\
0.7--0.8 & 61 & 0.749 & 0.787 & +0.038 \\
0.8--0.9 & 71 & 0.853 & 0.901 & +0.048 \\
0.9--1.0 & 125 & 0.951 & 0.912 & -0.039 \\
\midrule
\multicolumn{5}{l}{ECE 0.049 [0.034, 0.095], MCE 0.373, Brier 0.331, NLL 0.578} \\
\bottomrule
\end{tabular}
\caption{\rev{Statistics showing the accuracy of the model at different levels of confidence based on classifier output probabilities. In general, and according to the expected calibration error (ECE), the model is well calibrated.}}
\label{tab:stance-confidence}
\end{table*}


\begin{table*}[]
\centering
\begin{tabular}{lrrrrrrrrr}
\toprule
 &  & \multicolumn{3}{c}{Target extraction} & \multicolumn{5}{c}{Stance} \\
\cmidrule(lr){3-5} \cmidrule(lr){6-10}
Subgroup & Posts & $n$ & Error \% & 95\% CI & $n$ & Error \% & 95\% CI & Macro-F1 & $\kappa$ \\
\midrule
\multicolumn{10}{l}{\textit{platform}} \\
twitter & 73 & 128 & 12.5 & [7.8, 19.3] & 112 & 22.3 & [15.6, 30.9] & 0.755 & 0.644 \\
instagram & 75 & 127 & 15.7 & [10.4, 23.1] & 107 & 13.1 & [8.0, 20.8] & 0.792 & 0.718 \\
tiktok & 61 & 116 & 26.7 & [19.5, 35.4] & 85 & 38.8 & [29.2, 49.5] & 0.604 & 0.414 \\
bluesky & 41 & 73 & 2.7 & [0.8, 9.5] & 71 & 23.9 & [15.5, 35.0] & 0.753 & 0.630 \\
\midrule
\multicolumn{10}{l}{\textit{year}} \\
2024 & 54 & 98 & 16.3 & [10.3, 24.9] & 82 & 22.0 & [14.4, 32.1] & 0.760 & 0.655 \\
2023 & 54 & 93 & 9.7 & [5.2, 17.4] & 84 & 26.2 & [18.0, 36.5] & 0.703 & 0.567 \\
2025 & 52 & 93 & 14.0 & [8.4, 22.5] & 80 & 25.0 & [16.8, 35.5] & 0.650 & 0.573 \\
2026 & 49 & 87 & 18.4 & [11.6, 27.8] & 71 & 23.9 & [15.5, 35.0] & 0.760 & 0.630 \\
2022 & 41 & 73 & 20.5 & [12.9, 31.2] & 58 & 20.7 & [12.3, 32.8] & 0.767 & 0.653 \\
\midrule
\multicolumn{10}{l}{\textit{political group}} \\
Green & 25 & 46 & 19.6 & [10.7, 33.2] & 37 & 13.5 & [5.9, 28.0] & 0.830 & 0.778 \\
Bloc Québécois & 26 & 44 & 13.6 & [6.4, 26.7] & 38 & 34.2 & [21.2, 50.1] & 0.657 & 0.490 \\
Independent & 22 & 42 & 16.7 & [8.3, 30.6] & 35 & 17.1 & [8.1, 32.7] & 0.772 & 0.712 \\
Liberal & 22 & 40 & 10.0 & [4.0, 23.1] & 36 & 27.8 & [15.8, 44.0] & 0.651 & 0.464 \\
unaffiliated politician & 22 & 39 & 15.4 & [7.2, 29.7] & 33 & 18.2 & [8.6, 34.4] & 0.694 & 0.404 \\
Other party & 22 & 39 & 15.4 & [7.2, 29.7] & 33 & 24.2 & [12.8, 41.0] & 0.739 & 0.601 \\
Conservative & 22 & 38 & 10.5 & [4.2, 24.1] & 34 & 29.4 & [16.8, 46.2] & 0.587 & 0.478 \\
NDP & 20 & 35 & 22.9 & [12.1, 39.0] & 27 & 3.7 & [0.7, 18.3] & 0.961 & 0.922 \\
PPC\,$^{\dagger}$ & 10 & 18 & 5.6 & [1.0, 25.8] & 17 & 35.3 & [17.3, 58.7] & 0.566 & 0.333 \\
\midrule
\multicolumn{10}{l}{\textit{actor type}} \\
politician & 191 & 341 & 15.0 & [11.6, 19.1] & 290 & 22.4 & [18.0, 27.6] & 0.735 & 0.617 \\
influencer & 38 & 68 & 14.7 & [8.2, 25.0] & 58 & 24.1 & [15.0, 36.5] & 0.766 & 0.635 \\
foreign & 21 & 35 & 22.9 & [12.1, 39.0] & 27 & 37.0 & [21.5, 55.8] & 0.607 & 0.449 \\
\midrule
\multicolumn{10}{l}{\textit{text provenance}} \\
written & 188 & 326 & 12.3 & [9.1, 16.3] & 286 & 18.9 & [14.8, 23.8] & 0.783 & 0.689 \\
transcript (ASR) & 42 & 81 & 27.2 & [18.7, 37.7] & 59 & 47.5 & [35.3, 60.0] & 0.512 & 0.284 \\
image text (OCR) & 20 & 37 & 18.9 & [9.5, 34.2] & 30 & 23.3 & [11.8, 40.9] & 0.697 & 0.545 \\
\bottomrule
\end{tabular}
\caption{\rev{Subgroup error rates for stance target extraction and stance detection. A dagger indicates that a row has fewer than 20 pairs, and so rates should be treated with caution. We annotated 444 targets from 250 posts, which is why $n$ for Target extraction is higher than the number of posts. Stance error is only reported for targets judged relevant, which is why stance $n$ is lower than target extraction $n$.}}
\label{tab:subgroup-error-rates}
\end{table*}

\section{Latent-GP Factor Analysis}
\label{sec:lgpfa}

{\color{revcolor}%
The probability of the final stance is determined via an ordinal likelihood:%
}

{\color{revcolor}%
\begin{align}
  \Pr(y=-1\mid f) &= \Phi(-c-f),\\
  \Pr(y=0\mid f)  &= \Phi(c-f)-\Phi(-c-f),\\
  \Pr(y=+1\mid f) &= \Phi(f-c)
  \label{eq:probit}
\end{align}%
}

{\color{revcolor}%
Where $y$ is the stance of the post, and $c$ is a learned threshold. We represent the GP as a linear-Gaussian state-space model (SSM), to increase the tractability of training the model \citep{hartikainen2010kalman}.  We experimented with using an Ornstein-Uhlenbeck kernel but found that its mean-reversion property led to the prior dominating the landscape model characteristics in a way not entirely representative of the data, combined with the fact that intuitively, stance change is not mean-reverting.%
}

{\color{revcolor}%
We are using an ordinal likelihood with our GPFA, so we cannot use a closed-form expectation maximization process. We therefore seek a surrogate Gaussian to approximate the ordinal likelihood, in order to create a linear-Gaussian state-space model, for which we can use an expectation maximization process, where the expectation step is a Kalman filter plus a Rauch–Tung–Striebel (RTS) smoother. To do this, we use conjugate-computation variational inference \citep{khan2017conjugate}.%
}

{\color{revcolor}%
The learned thresholds $(-c, c)$ are symmetric.%
}

{\color{revcolor}%
At a given point, we have the output predictor of the GPFA for each figure/target, $f_i$, which we use to predict the stance mixture results for a given cell, i.e. $\log p(\text{cell
} \mid f)$. We trial three methods of operationalizing that probability, given that the stance classification model produces a label prediction, probability outputs, and we know the classifier's performance on the test-set at train time:%
}

{\color{revcolor}%
\begin{itemize}
    \item \textbf{Hard}: Consider only the classifier's final label, and aggregate them for a given cell: $\log p = n_{against}\log P(-1\mid f) + n_{neutral}\log P(0\mid f) + n_{favor}\log P(+1\mid f)$
    \item \textbf{Channel}: Consider the classifier's final label, in addition to the test confusion matrix.
    \item \textbf{Soft}: Consider only the classifier's final probabilities, so the cell's sufficient statistics are the sum of the probabilities for each label from all of the cell's post-target pairs: $\sum_iq_{ik}$.
\end{itemize}%
}

{\color{revcolor}%
We show the swept parameter ranges in Table \ref{tab:lgpfa-sweep-parameters}, and we list the final parameters used when fitting the latent-GP factor analysis model in Table \ref{tab:lgpfa-parameters}. %
}

\begin{table}[]
    \centering
    \begin{tabular}{c|c}
    \toprule
        Parameter & Range \\
    \midrule
        N. Dims & [3, 4, 5, 6] \\
        N. Fast Dims & [1, 2, 3, 4, 6] \\
        Tau & [10, 20, 40, 80] \\
        Slow Kernel & [Const, Wiener] \\
        Slow Tau & [640, 2560] \\
        Observation Model & [Hard, Soft, Channel] \\
        Observation Temperature & [1, 1.33, 1.8, 2.5, 3.4] \\
        Loadings Ridge ($w_{ridge}$) & [0.0001, 1, 10, 100] \\
        \bottomrule    
        \end{tabular}    
        \caption{\rev{Parameter ranges used when fitting the latent-GP factor analysis model via hyperparameter sweeps. Note that the number of fast dimensions can be higher than the number of dimensions, in which case the lower number of dimensions will be used, and they will all be fast.}}
    \label{tab:lgpfa-sweep-parameters}
\end{table}

\begin{table}[]
    \centering
    \begin{tabular}{c|c}
    \toprule
        Parameter & Value \\
    \midrule
        N. Dims & 5 \\
        N. Fast Dims & 5 \\
        Kernel & Wiener \\
        Tau & 10 \\
        Slow Kernel & Wiener \\
        Slow Tau & 2560 \\
        Bin Factor & 8 \\
        Observation Model & Hard \\
        $\rho$ & 0 \\
        Loadings Ridge ($w_{ridge}$) & 1 \\
        EM Iters & 25 \\
        Infer Iters & 15 \\
        Min. Target Volume & 400 \\
        Causal State & True \\
        Seed & 0 \\
    \bottomrule
    \end{tabular}
    \caption{\rev{Parameters used when fitting the latent-GP factor analysis model, as determined through hyperparameter optimization.}}
    \label{tab:lgpfa-parameters}
\end{table}

{\color{revcolor}%
We tested the stability of the low-dimensional space the GPFA learns by re-training the model on two sets of disjoint people from the dataset, with this experiment repeated five times. We found that the first three dimensions are consistent, with a cosine of the principal angles between subspaces of more than 0.9 when weighting targets by post volume (where a null re-assignment of target loadings had a root-mean-square cosine of 0.3 averaged over the five dimensions).%
}

\begin{figure}
    \centering
    \includegraphics[width=0.9\linewidth]{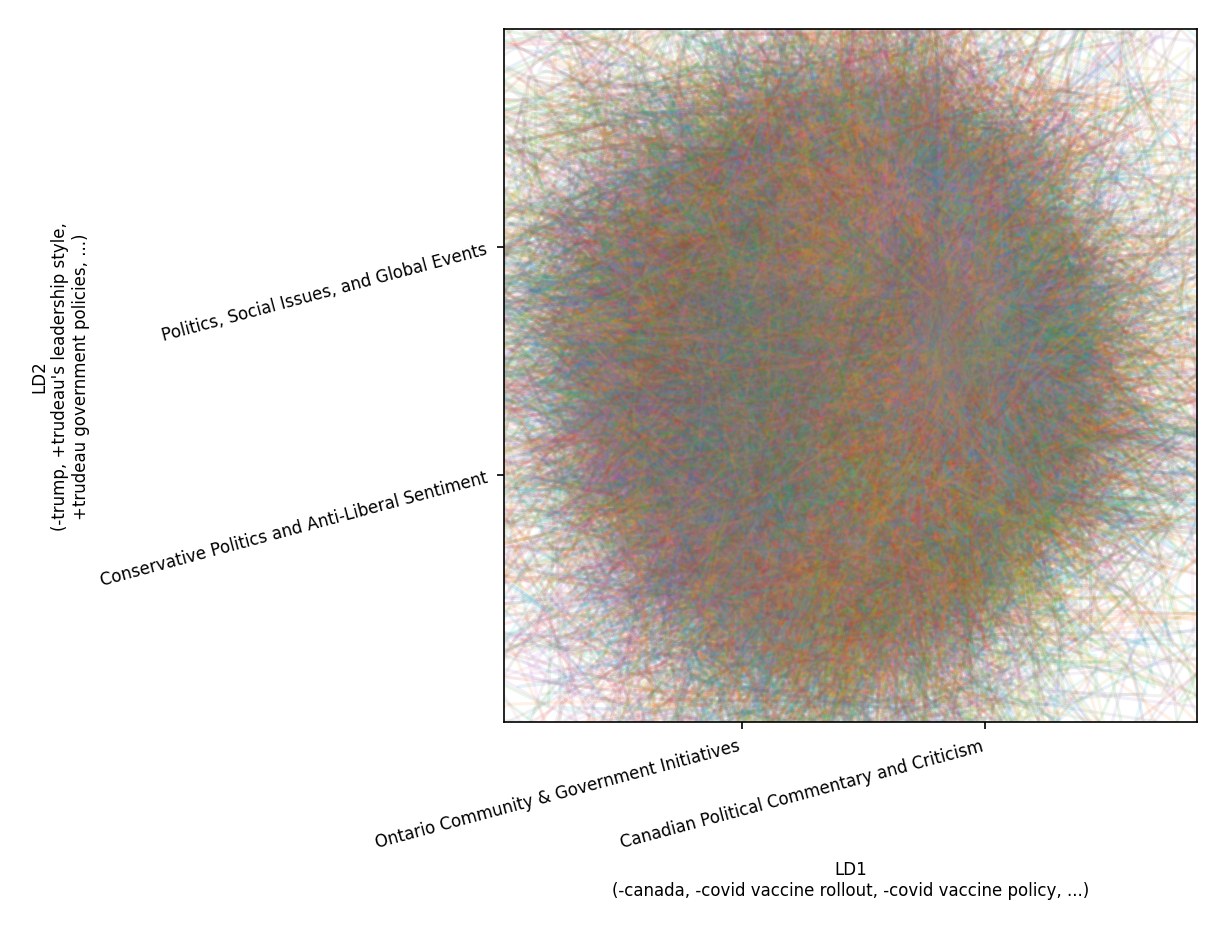}
    \caption{\rev{The first two dimensions of the latent stance space, shown with the raw trajectories, which are hard to interpret on their own.}}
    \label{fig:trajectories}
\end{figure}

{\color{revcolor}%
We include a figure of the raw trajectories from the GPFA step in Fig \ref{fig:trajectories}.%
}

\section{Stationarity}
\label{sec:app-station}

{\color{revcolor}%
We use Pesaran's CADF/CIPS test \citep{pesaran2007simple} (as the trajectories produced by the GPFA are not independent) and the Kwiatkowski-Phillips-Schmidt-Shin (KPSS) test \citep{kwiatkowski1992testing} on the time-varying trajectories from the GPFA. We see that CIPS produces a test statistic of -4.916 averaged over the first three dimensions, rejecting the unit root on every one of them ($p \leq 0.005$), and that at least 13\% of trajectories reject stationarity via the KPSS on every dimension ($p \leq 0.015$).%
}

\section{Markov assumption}
\label{sec:inertia}

One way to think about this design decision is to imagine two groups of people, one group of which had previously all held very strong positions on a subject, but both groups hold the same attitude at the current moment. They are then subject to the same persuasive content. Do the groups change their attitudes in response to the information in the same way? Or is there a path dependency on the causal effect of the information? \citet{coppock2023persuasion,coppock2020small,guess2020does} suggest that we should assume they update in the same way for mass publics, and \citet{feddersen2023changing} suggests that political elites are also able to do this irrespective of their previous positions. If there is no path dependency in the way their attitudes update, that implies a Markov assumption \textemdash that is, that the way they could move is dependent only on their current state, not on their previous states. This means we will use the simplifying assumption of not modelling inertia in the system. However, the evidence on this is mixed \citep{howe2017attitude}, and therefore this assumption should be relaxed in future if it produces a more predictive model.

\section{Landscape Model}
\label{sec:landscape-model}

{\color{revcolor}%
We train the MLP via JAX autograd, as in \citet{howe2025learning}. The time parameter is determined by normalising time-spans such that $1^{st}$ January 2022 is mapped to 1, and $1^{st}$ January 2025 is mapped to 4, and this continuous number is concatenated to the initial trajectory location in the MLP input vector, to produce an input vector of shape $1 + N_{coord dims}$. The output of the MLP is a scalar, the height of the potential landscape at that point.%
}

{\color{revcolor}%
In order to display two dimensional projections of the potential landscape, we marginalize over the other dimensions using the Boltzmann-weighted average \citep{roux1995calculation}.%
}

{\color{revcolor}%
We alter the confinement term from \citet{howe2025learning} to maintain its ability to confine trajectories to finite values, but reduce its influence on the periphery of the landscape. We therefore change it from $C_0||x'_{t+1}||^4$, to: $C_0 max(0, ||x'_{t+1}|| - r)^4$, where $r = 1.1 \cdot \max_{t, i} \lVert x_t^{(i)} \rVert$, and $C_0$, the confinement term, is a hyperparameter to optimize. This results in a full loss function of $||x_{t+1}'-x_{t+1}||_2 + C_0 max(0, ||x'_{t+1}||-r)^4$.%
}

{\color{revcolor}%
While using the MSE loss function sends a more direct gradient signal, it does mean that if the sigma term is trainable, the optimizer drives it to zero. %
}

{\color{revcolor}%
We derive the hyperparameter optimization objective of the sweeps in the following way:%
}

{\color{revcolor}%
Let each evaluation pair consist of a trajectory observed at $t_0$ and at
$t_1 = t_0 + h$, where $d = z(t_1) - z(t_0)$ is the observed displacement,
$p = \hat{z}(t_1) - z(t_0)$ is the displacement predicted by the model, and $m = z(t_{-1}) - z(t_0)$ is a simple reversion baseline model that we want to out-perform, in addition to the no movement baseline.
Let $D = \mathbf{E}\lVert d\rVert^2$, $P = \mathbf{E}\lVert p\rVert^2$,
$M = \mathbf{E}\lVert m\rVert^2$, and $C_{dp}, C_{dm}, C_{pm}$ the corresponding
cross-moments, all averaged over pairs.%
}

{\color{revcolor}%
Since the no-movement baseline's error is exactly $D$, the skill score
decomposes without residual as
\begin{align}
  S &= 1 - \frac{\mathbf{E}\lVert d - p\rVert^2}{\mathbf{E}\lVert d\rVert^2}\\
  \label{eq:skill}
  S &= 2\rho R - R^2\\
  \rho &= \frac{C_{dp}}{\sqrt{DP}}\\
  R &= \sqrt{P/D}
  \label{eq:r-def}
\end{align}
into a direction correlation $\rho$ and an amplitude ratio $R$. Maximising
\eqref{eq:skill} over rescalings of $p$ gives $\rho^2$, attained at $R = \rho$:
corresponding to the best skill that drift field could reach with any rescaling. We optimise
against $\rho^2$ rather than $S$ because $S \to 0$ as $p \to 0$ whatever the
model has learned, so it cannot distinguish a collapsed landscape from a poorly fit one.%
}

{\color{revcolor}%
The reversion rule reaches $\lvert\rho_{dm}\rvert \approx 0.25$ on these latents
unaided, so $\rho^2$ alone would credit the landscape for rediscovering it.
With $\rho_{dm} = C_{dm}/\sqrt{DM}$ and $\rho_{pm} = C_{pm}/\sqrt{PM}$, the
variance of the observed motion explained by both predictors together is
\begin{equation}
  R^2_{\mathrm{comb}} =
  \frac{\rho^2 + \rho_{dm}^2 - 2\rho\,\rho_{dm}\,\rho_{pm}}{1 - \rho_{pm}^2},
  \label{eq:combined}
\end{equation}
clipped to $[0,1]$, and the sweep maximises what the landscape adds beyond
reversion,
\begin{equation}
  \mathrm{gain} = R^2_{\mathrm{comb}} - \rho_{dm}^2,
  \label{eq:gain}
\end{equation}
evaluated at a 96-day horizon on validation trajectories in the held-out time
span.%
}

{\color{revcolor}%
We used an Euler solver for the integrator of the potential landscape model.%
}

\rev{We show the hyperparameter ranges used for hyperparameter sweeps in Table \ref{tab:potential-sweep-parameters}.} We list the hyperparameters used in our potential landscape model in Table \ref{tab:potential-model-hyperparameters}. We use the softplus activation function as in \citet{howe2025learning}.

\begin{table}[]
    \centering
    \begin{tabular}{c|c}
    \toprule
        Parameter & Value \\
    \midrule
        N. Dims & [3, 4, 5, 6] \\
        $\sigma_{initial}$ & $[0.02, 0.4]$, log-uniform \\
        Hidden Dims & [$64^3$, $128^4$, $256^3$] \\
        Dropout & $[0, 0.2]$, uniform \\
        Weight Decay & $[10^{-4}, 10^{-1}]$, log-uniform \\
        Learning Rate & $[10^{-4}, 10^{-2}]$, log-uniform \\
        Confinement Factor & $[10^{-4}, 10^{-1}]$, log-uniform \\
    \bottomrule
    \end{tabular}    
    \caption{\rev{Parameter ranges explored for the landscape model when doing hyperparameter sweeps.}}
    \label{tab:potential-sweep-parameters}
\end{table}

\begin{table}[]
    \centering
    \begin{tabular}{c|c}
    \toprule
        Hyperparameter & Value \\
    \midrule
        N. Dims & 5 \\
        Batch Size & 1024 \\
        Num Epochs & 100 \\
        Patience & 10 \\
        Train Fraction & 0.7 \\
        Val Fraction & 0.1 \\
        $\sigma_{initial}$ & 0.029 \\
        Hidden Dims & [64, 64, 64] \\
        Dropout & 0.063 \\
        Weight Decay & 0.049 \\
        Learning Rate & 0.00083 \\
        Confinement Factor & 0.011 \\
    \bottomrule
    \end{tabular}
    \caption{\rev{Hyperparameters used for the potential landscape model, with some parameters determined through hyperparameter optimization.}}
    \label{tab:potential-model-hyperparameters}
\end{table}

When training the landscape models on \rev{platform-specific} data, we found that increasing the dropout to 0.3, and the weight decay to 0.05, helped to improve predictive performance in this case due to the reduced quantity of data. We trialled \rev{fine-tuning the platform-specific} models from the general model, but this did not improve predictive performance on \rev{platform-specific} data.

{\color{revcolor}%
We additionally compared our landscape model to a Holt-damped trend forecasting method, which we found to be worse than both the damped-Theta method and the mean-reverting method, so we dropped it.%
}

\section{Dimensional Analysis}
\label{sec:dim-analysis}

{\color{revcolor}%
We additionally see that in the fourth and fifth latent dimensions, in the rankings of the party pairs with the furthest centroids, the Bloc Québécois party dominates the highest ranked pair distances. The Bloc Québécois is a political party focused on furthering Québécois interests, a distinct French-language constituency in Canada, with heterodox political positions.  %
}

\section{Survey Results}
\label{sec:survey-results}

We compare the stance movement we see to equivalent measures from \rev{an ongoing} longitudinal survey tracking attitudes in Canada, ranging from 2024 through 2025 \citep{bridgman2023canadian}. \rev{However, these results come from members of the general public rather than political elites as in this work, so the extent to which we can compare the two results is limited.} There is no significant movement in average self-reported left-right ideology between August 2024 and December 2025. There is also no overall trend in trust in elected officials, the police, or journalists when measured from `No trust' to `A lot of trust'. This disparity may be explained by self-reports \textemdash people don't perceive that their ideology has changed when placed on a 1-10 number line, but their expressed stance has changed on some issues, aligning with prior work \citep{ellis2012public}. It could also be explained by expressed attitude not aligning with private attitudes. However, we do see positive movement on the Liberal party in 2025, with Mark Carney's win of the leadership race and election resulting in a large increase in points on a feeling thermometer question in the survey.

From surveys of Canadians, we know that people whose most-used platform is Instagram or Bluesky generally place the Liberal party highest on a feeling thermometer compared to other Canadian parties \citep{bridgman2023canadian}. So the relative \rev{higher position of the attractors for Instagram and Bluesky} on the second \rev{dimension make} sense. Conversely, in general, the party rated highest on a \rev{feeling} thermometer of those who say X/Twitter is their most-used platform is the Conservative party. \rev{Therefore, the attractor being lower} on the second \rev{LD makes sense for} Twitter. TikTok has no clear ordering of political party by those who cite it as their most-used platform.

\section{Stance Change Examples}
\label{sec:stance-examples}

We use the $\chi^2$ test for cases where there are posts for all three stance labels, and \rev{Fisher's exact} test for cases where there are posts on two stance labels.

In Table \ref{tab:all-dims-examples}, we show the top three significant ($p < 0.05$) examples of movement for different \rev{percentile} movers on each dimension.

\begin{table*}[]
    \centering
    \begin{tabular}{llllrrr}
    \toprule
    Dim. & Dir. & Percentile & Target & Favor & Neutral & Against \\
    \midrule
    LD1 & Pos. & Top 10\% & climate change policy & 46.3→31.0 & 38.6→44.4 & 15.0→24.6 \\
     &  &  & transgender rights & 58.8→41.0 & 30.5→38.5 & 10.7→20.5 \\
     &  &  & gender identity protections & 49.1→31.3 & 39.6→48.6 & 11.3→20.1 \\
    \midrule
    LD1 & Pos. & Top 1\% & canada & 49.9→34.2 & 31.9→45.1 & 18.2→20.7 \\
     &  &  & children & 72.7→42.4 & 25.0→52.1 & 2.3→5.6 \\
    \midrule
    LD1 & Neg. & Top 10\% & student academic performance & 37.4→44.7 & 60.7→52.5 & 1.9→2.8 \\
     &  &  & community & 91.8→95.5 & 7.8→4.3 & 0.3→0.2 \\
    \midrule
    LD1 & Neg. & Top 1\% & voting rights & 41.0→64.5 & 58.2→34.1 & 0.8→1.4 \\
    \midrule
    LD2 & Pos. & Top 10\% & donald trump & 57.2→36.2 & 14.2→21.1 & 28.6→42.7 \\
     &  &  & conservatives & 44.8→24.1 & 34.7→43.5 & 20.4→32.3 \\
     &  &  & trump & 34.3→23.6 & 24.2→34.1 & 41.6→42.4 \\
    \midrule
    LD2 & Pos. & Top 1\% & trump & 46.2→15.5 & 20.4→25.9 & 33.3→58.6 \\
     &  &  & donald trump & 65.9→29.3 & 9.8→19.5 & 24.4→51.2 \\
     &  &  & trudeau government policies & 1.2→12.7 & 22.3→43.1 & 76.5→44.2 \\
    \midrule
    LD2 & Neg. & Top 10\% & indigenous immigration policies & 29.3→17.9 & 55.6→47.9 & 15.1→34.2 \\
     &  &  & conservatives & 36.5→47.3 & 35.8→29.7 & 27.7→23.0 \\
     &  &  & climate change policy & 43.5→30.6 & 41.4→49.1 & 15.1→20.4 \\
    \midrule
    LD2 & Neg. & Top 1\% & mark carney & 45.9→15.9 & 35.1→41.4 & 18.9→42.7 \\
     &  &  & carbon emissions regulation & 65.9→36.8 & 28.8→59.2 & 5.3→3.9 \\
     &  &  & climate change policy & 70.1→43.9 & 21.2→50.5 & 8.7→5.5 \\
    \midrule
    LD3 & Pos. & Top 10\% & mark carney's economic policies & 18.8→4.4 & 48.0→33.9 & 33.2→61.8 \\
     &  &  & proposed bills in canada & 40.2→22.0 & 20.5→19.7 & 39.3→58.3 \\
     &  &  & mark carney & 30.4→9.3 & 43.0→35.1 & 26.6→55.6 \\
    \midrule
    LD3 & Pos. & Top 1\% & liberal party policies & 6.0→1.6 & 38.9→14.9 & 55.1→83.5 \\
     &  &  & liberal government approach & 5.4→0.9 & 38.9→16.5 & 55.7→82.6 \\
     &  &  & trudeau's cabinet decisions & 2.5→3.7 & 60.0→32.4 & 37.4→63.9 \\
    \midrule
    LD3 & Neg. & Top 10\% & joe biden & 13.9→10.5 & 31.1→62.6 & 55.1→26.9 \\
     &  &  & joe biden's policies & 9.4→6.9 & 37.6→69.2 & 53.0→23.9 \\
     &  &  & biden & 12.8→12.1 & 33.2→45.0 & 54.1→42.9 \\
    \midrule
    LD3 & Neg. & Top 1\% & joe biden's policies & 9.8→9.7 & 43.8→67.7 & 46.5→22.6 \\
     &  &  & joe biden & 15.1→13.7 & 36.0→59.3 & 48.9→26.9 \\
     &  &  & carbon tax implementation & 13.4→10.3 & 15.0→39.7 & 71.7→50.0 \\
    \bottomrule
    \end{tabular}
    \caption{\rev{Examples of significant movement across all time for each of the first three latent dimensions. Numbers/arrows indicate old and new percentage over time.}}
    \label{tab:all-dims-examples}
\end{table*}

In Table \ref{tab:2022-examples} we include the top three significant ($p < 0.05$) examples of stance changes on targets for 2022 from our dataset.

\begin{table*}[]
    \centering
    \begin{tabular}{llllrrr}
    \toprule
    Dim. & Dir. & Percentile & Target & Favor & Neutral & Against \\
    \midrule
    LD1 & Pos. & Top 10\% & canada & 38.1→24.7 & 44.0→47.5 & 17.8→27.8 \\
     &  &  & leadership & 43.1→22.4 & 29.2→10.3 & 27.7→67.3 \\
     &  &  & jews & 39.1→16.5 & 45.7→42.4 & 15.2→41.2 \\
    \midrule
    LD1 & Neg. & Top 10\% & faith-based worldviews & 42.9→86.8 & 36.5→5.3 & 20.6→7.9 \\
     &  &  & religious belief systems & 42.9→84.2 & 39.7→13.2 & 17.5→2.6 \\
     &  &  & russian foreign policy & 2.3→0.0 & 63.0→35.2 & 34.8→64.8 \\
    \midrule
    LD2 & Pos. & Top 10\% & trump & 78.7→71.7 & 9.1→12.8 & 12.3→15.5 \\
     &  &  & climate change policy & 53.6→61.3 & 34.1→32.4 & 12.3→6.3 \\
     &  &  & trudeau government policies & 2.6→3.9 & 34.1→37.2 & 63.3→58.9 \\
    \midrule
    LD2 & Pos. & Top 1\% & joe biden's policies & 0.0→50.0 & 47.1→16.7 & 52.9→33.3 \\
     &  &  & joe biden & 0.0→50.0 & 47.1→8.3 & 52.9→41.7 \\
    \midrule
    LD2 & Neg. & Top 10\% & israel & 13.3→52.6 & 53.3→42.1 & 33.3→5.3 \\
     &  &  & trump & 11.0→18.4 & 27.5→32.2 & 61.4→49.3 \\
     &  &  & faith-based worldviews & 52.0→77.1 & 22.0→12.0 & 26.0→10.8 \\
    \midrule
    LD2 & Neg. & Top 1\% & conservatives & 64.0→95.0 & 12.0→5.0 & 24.0→0.0 \\
     &  &  & covid vaccine rollout & 12.2→9.1 & 45.2→18.2 & 42.6→72.7 \\
     &  &  & freedom of speech & 59.0→88.5 & 41.0→11.5 & 0.0→0.0 \\
    \bottomrule
    \end{tabular}
    \caption{\rev{Examples of significant stance change on stance targets that are loaded on by the first two latent dimensions, in 2022. Numbers/arrows indicate old and new percentage over time.}}
    \label{tab:2022-examples}
\end{table*}

In Table \ref{tab:2023-examples} we include the top three significant ($p < 0.05$) examples of stance changes on targets for 2023 from our dataset.

\begin{table*}[]
    \centering
    \begin{tabular}{llllrrr}
    \toprule
    Dim. & Dir. & Percentile & Target & Favor & Neutral & Against \\
    \midrule
    LD1 & Pos. & Top 10\% & indigenous immigration policies & 29.0→13.2 & 48.5→46.7 & 22.5→40.1 \\
     &  &  & vote & 70.8→34.0 & 22.0→46.9 & 7.1→19.0 \\
     &  &  & immigration & 25.6→9.5 & 35.1→35.7 & 39.3→54.9 \\
    \midrule
    LD1 & Pos. & Top 1\% & canada & 64.3→14.8 & 21.4→46.6 & 14.3→38.6 \\
    \midrule
    LD1 & Neg. & Top 10\% & trudeau's leadership style & 4.2→2.5 & 28.5→46.5 & 67.2→50.9 \\
     &  &  & proposed bills in canada & 25.7→38.7 & 20.1→19.1 & 54.3→42.2 \\
     &  &  & trudeau government policies & 2.6→1.6 & 31.7→49.4 & 65.7→49.0 \\
    \midrule
    LD2 & Pos. & Top 10\% & bc conservatives & 75.0→28.8 & 8.3→33.3 & 16.7→37.9 \\
     &  &  & conservatives & 32.9→19.8 & 49.5→59.6 & 17.6→20.6 \\
     &  &  & trudeau's leadership style & 4.1→5.4 & 29.4→36.5 & 66.5→58.2 \\
    \midrule
    LD2 & Pos. & Top 1\% & trump & 83.3→28.0 & 11.1→32.0 & 5.6→40.0 \\
     &  &  & america & 70.0→6.2 & 20.0→81.2 & 10.0→12.5 \\
     &  &  & american diplomacy & 50.0→4.6 & 50.0→86.2 & 0.0→9.2 \\
    \midrule
    LD2 & Neg. & Top 10\% & indigenous immigration policies & 35.0→23.7 & 56.9→54.7 & 8.2→21.6 \\
     &  &  & vaccine & 36.0→22.3 & 38.7→35.1 & 25.3→42.6 \\
     &  &  & carbon pricing mechanisms & 12.7→7.5 & 38.4→30.8 & 48.9→61.7 \\
    \bottomrule
    \end{tabular}
    \caption{\rev{Examples of significant stance change on stance targets that are loaded on by the first two latent dimensions, in 2023. Numbers/arrows indicate old and new percentage over time.}}
    \label{tab:2023-examples}
\end{table*}

In Table \ref{tab:2024-examples} we include the top three significant ($p < 0.05$) examples of stance changes on targets for 2024 from our dataset.

\begin{table*}[]
    \centering
    \begin{tabular}{llllrrr}
    \toprule
    Dim. & Dir. & Percentile & Target & Favor & Neutral & Against \\
    \midrule
    LD1 & Pos. & Top 10\% & kamala harris & 90.4→56.8 & 3.8→20.3 & 5.7→22.9 \\
     &  &  & proposed bills in canada & 43.4→12.5 & 21.5→27.2 & 35.2→60.3 \\
     &  &  & kamala & 65.1→42.1 & 23.3→27.9 & 11.6→30.0 \\
    \midrule
    LD1 & Pos. & Top 1\% & canada & 59.7→32.8 & 28.4→49.3 & 11.9→17.9 \\
     &  &  & government & 27.7→11.9 & 42.6→34.6 & 29.8→53.5 \\
     &  &  & election & 46.5→26.1 & 53.5→69.7 & 0.0→4.1 \\
    \midrule
    LD1 & Neg. & Top 10\% & canada & 23.4→32.8 & 51.7→45.1 & 24.9→22.1 \\
     &  &  & teachers & 47.2→64.3 & 28.3→27.4 & 24.5→8.3 \\
     &  &  & liberal government approach & 3.1→4.9 & 25.5→29.1 & 71.4→66.0 \\
    \midrule
    LD1 & Neg. & Top 1\% & trump & 42.7→28.2 & 36.2→46.2 & 21.2→25.6 \\
    \midrule
    LD2 & Pos. & Top 10\% & trump & 46.3→34.0 & 27.6→33.2 & 26.1→32.8 \\
     &  &  & donald trump & 59.0→48.6 & 15.3→19.6 & 25.7→31.8 \\
     &  &  & trump's political agenda & 63.8→41.5 & 23.4→40.8 & 12.8→17.7 \\
    \midrule
    LD2 & Pos. & Top 1\% & pierre poilievre & 53.2→43.0 & 30.4→27.4 & 16.5→29.7 \\
     &  &  & bc conservatives & 64.3→31.6 & 19.0→28.1 & 16.7→40.4 \\
     &  &  & carbon tax implementation & 3.1→11.9 & 23.9→37.6 & 73.0→50.5 \\
    \midrule
    LD2 & Neg. & Top 10\% & joe biden & 32.4→17.8 & 39.7→53.7 & 27.9→28.5 \\
     &  &  & biden & 31.4→21.0 & 41.4→39.0 & 27.2→40.1 \\
     &  &  & conservatives & 21.2→32.4 & 49.3→41.2 & 29.5→26.4 \\
    \midrule
    LD2 & Neg. & Top 1\% & israel & 2.2→23.4 & 42.2→30.7 & 55.6→45.9 \\
     &  &  & palestinian & 46.2→26.3 & 42.8→26.3 & 11.0→47.4 \\
     &  &  & jewish community advocacy & 25.3→58.9 & 57.8→32.9 & 16.9→8.2 \\
    \bottomrule
    \end{tabular}
    \caption{\rev{Examples of significant stance change on stance targets that are loaded on by the first two latent dimensions, in 2024. Numbers/arrows indicate old and new percentage over time.}}
    \label{tab:2024-examples}
\end{table*}

In Table \ref{tab:2025-examples} we include the top three significant ($p < 0.05$) examples of stance changes on targets for 2025 from our dataset.

\begin{table*}[]
    \centering
    \begin{tabular}{llllrrr}
    \toprule
    Dim. & Dir. & Percentile & Target & Favor & Neutral & Against \\
    \midrule
    LD1 & Pos. & Top 10\% & canada & 43.7→33.0 & 40.0→44.0 & 16.3→23.0 \\
     &  &  & transgender rights & 55.3→33.7 & 36.2→48.7 & 8.5→17.7 \\
     &  &  & indigenous immigration policies & 24.5→15.6 & 55.4→51.2 & 20.0→33.3 \\
    \midrule
    LD1 & Pos. & Top 1\% & canada & 58.1→39.5 & 30.4→42.9 & 11.4→17.6 \\
    \midrule
    LD1 & Neg. & Top 10\% & trudeau's leadership style & 6.7→3.5 & 27.1→39.9 & 66.1→56.6 \\
     &  &  & trudeau's cabinet decisions & 4.7→2.5 & 31.1→42.7 & 64.2→54.8 \\
    \midrule
    LD2 & Pos. & Top 10\% & faith-based worldviews & 56.2→42.0 & 29.9→37.4 & 13.9→20.6 \\
     &  &  & donald trump & 31.6→24.8 & 19.6→22.1 & 48.9→53.1 \\
     &  &  & conservative government & 88.9→70.9 & 3.8→10.6 & 7.3→18.4 \\
    \midrule
    LD2 & Neg. & Top 10\% & palestinian & 62.8→48.9 & 25.1→27.2 & 12.1→23.8 \\
     &  &  & palestine & 64.0→56.2 & 18.8→23.0 & 17.2→20.8 \\
     &  &  & indigenous immigration policies & 21.9→13.4 & 49.3→53.3 & 28.8→33.3 \\
    \midrule
    LD2 & Neg. & Top 1\% & trudeau's leadership style & 28.7→3.8 & 45.7→52.9 & 25.6→43.3 \\
     &  &  & justin trudeau & 47.5→11.3 & 35.5→47.9 & 17.0→40.8 \\
     &  &  & trudeau government policies & 22.1→1.9 & 52.5→57.7 & 25.4→40.4 \\
    \bottomrule
    \end{tabular}
    \caption{\rev{Examples of significant stance change on stance targets that are loaded on by the first two latent dimensions, in 2025. Numbers/arrows indicate old and new percentage over time.}}
    \label{tab:2025-examples}
\end{table*}

In Table \ref{tab:twitter-examples} we include the top three significant ($p < 0.05$) examples of stance changes on targets for Twitter/X from our dataset.

\begin{table*}[]
    \centering
    \begin{tabular}{llllrrr}
    \toprule
    Dim. & Dir. & Percentile & Target & Favor & Neutral & Against \\
    \midrule
    LD1 & Pos. & Top 10\% & transgender rights & 43.0→24.7 & 42.9→47.2 & 14.1→28.1 \\
     &  &  & covid vaccine policy & 19.8→12.7 & 59.6→49.3 & 20.6→38.1 \\
     &  &  & covid vaccine rollout & 23.0→15.4 & 58.9→50.4 & 18.2→34.2 \\
    \midrule
    LD1 & Pos. & Top 1\% & government & 27.5→14.7 & 26.9→29.7 & 45.5→55.6 \\
     &  &  & canada & 31.7→19.8 & 43.6→53.6 & 24.7→26.7 \\
    \midrule
    LD1 & Neg. & Top 10\% & kamala harris & 18.0→41.0 & 28.1→33.3 & 53.9→25.7 \\
     &  &  & kamala & 14.9→24.7 & 26.6→42.7 & 58.5→32.6 \\
     &  &  & carney & 3.2→6.2 & 33.0→46.7 & 63.8→47.1 \\
    \midrule
    LD2 & Pos. & Top 10\% & donald trump & 55.6→33.5 & 15.0→23.9 & 29.3→42.7 \\
     &  &  & trump & 38.3→25.9 & 28.8→37.4 & 33.0→36.7 \\
     &  &  & maga & 41.7→13.5 & 36.6→47.6 & 21.7→38.9 \\
    \midrule
    LD2 & Pos. & Top 1\% & trump & 34.7→12.3 & 36.0→44.5 & 29.3→43.2 \\
     &  &  & israel & 44.8→1.7 & 31.0→39.9 & 24.1→58.3 \\
     &  &  & elon musk & 55.6→13.2 & 33.3→34.0 & 11.1→52.8 \\
    \midrule
    LD2 & Neg. & Top 10\% & conservatives & 31.9→48.3 & 40.0→29.5 & 28.1→22.2 \\
     &  &  & ndp & 15.5→9.2 & 31.2→19.0 & 53.3→71.8 \\
     &  &  & democrats & 12.7→9.7 & 48.6→30.7 & 38.8→59.6 \\
    \midrule
    LD2 & Neg. & Top 1\% & faith-based worldviews & 8.3→36.5 & 0.0→46.0 & 91.7→17.5 \\
     &  &  & religious belief systems & 8.3→36.5 & 16.7→54.0 & 75.0→9.5 \\
     &  &  & biden & 80.0→46.8 & 12.5→33.1 & 7.5→20.2 \\
    \bottomrule
    \end{tabular}
    \caption{\rev{Examples of significant stance change on stance targets that are loaded on by the first two latent dimensions, for Twitter/X. Numbers/arrows indicate old and new percentage over time.}}
    \label{tab:twitter-examples}
\end{table*}

In Table \rev{\ref{tab:tiktok-examples}} we include the top three significant ($p < 0.05$) examples of stance changes on targets for \rev{TikTok} from our dataset.

\begin{table*}[]
    \centering
    \begin{tabular}{llllrrr}
    \toprule
    Dim. & Dir. & Percentile & Target & Favor & Neutral & Against \\
    \midrule
    LD1 & Pos. & Top 10\% & mark carney & 71.4→19.4 & 28.6→30.6 & 0.0→50.0 \\
     &  &  & mark carney's economic policies & 42.7→11.8 & 53.1→33.6 & 4.2→54.6 \\
     &  &  & climate change policy & 64.3→37.2 & 28.6→40.4 & 7.1→22.3 \\
    \midrule
    LD1 & Neg. & Top 10\% & trump & 46.0→13.8 & 13.7→10.0 & 40.3→76.2 \\
     &  &  & joe biden's policies & 26.7→28.9 & 35.6→61.8 & 37.8→9.2 \\
     &  &  & joe biden & 32.2→33.6 & 31.1→56.6 & 36.7→9.9 \\
    \midrule
    LD2 & Pos. & Top 10\% & trump & 40.0→27.0 & 12.2→17.2 & 47.8→55.8 \\
     &  &  & donald trump & 36.0→16.4 & 4.0→0.8 & 60.0→82.8 \\
     &  &  & joe biden & 32.8→27.8 & 21.1→54.4 & 46.1→17.8 \\
    \midrule
    LD2 & Neg. & Top 10\% & israel & 6.5→15.3 & 5.4→16.5 & 88.2→68.2 \\
     &  &  & climate change policy & 51.1→31.6 & 25.9→33.7 & 23.0→34.7 \\
     &  &  & mark carney & 16.7→6.6 & 26.7→12.1 & 56.7→81.3 \\
    \bottomrule
    \end{tabular}
    \caption{\rev{Examples of significant stance change on stance targets that are loaded on by the first two latent dimensions, for TikTok. Numbers/arrows indicate old and new percentage over time.}}
    \label{tab:tiktok-examples}
\end{table*}

In Table \ref{tab:instagram-examples} we include the top three significant ($p < 0.05$) examples of stance changes on targets for Instagram from our dataset.

\begin{table*}[]
    \centering
    \begin{tabular}{llllrrr}
    \toprule
    Dim. & Dir. & Percentile & Target & Favor & Neutral & Against \\
    \midrule
    LD1 & Pos. & Top 10\% & abortion & 68.8→11.5 & 0.0→15.4 & 31.2→73.1 \\
     &  &  & canada & 65.1→49.8 & 27.1→33.8 & 7.8→16.4 \\
     &  &  & mark carney & 58.2→15.3 & 18.2→36.4 & 23.6→48.3 \\
    \midrule
    LD1 & Pos. & Top 1\% & government & 75.0→31.8 & 16.7→22.7 & 8.3→45.5 \\
     &  &  & canada & 81.4→74.4 & 18.0→13.9 & 0.6→11.7 \\
     &  &  & liberal media & 9.1→0.0 & 54.5→16.1 & 36.4→83.9 \\
    \midrule
    LD2 & Pos. & Top 10\% & donald trump & 59.8→39.5 & 1.1→2.6 & 39.1→57.9 \\
     &  &  & trudeau's leadership style & 2.0→6.6 & 9.0→16.9 & 89.0→76.5 \\
     &  &  & trudeau government policies & 1.5→5.2 & 10.5→20.2 & 88.1→74.6 \\
    \midrule
    LD2 & Pos. & Top 1\% & trump & 57.6→10.3 & 12.1→7.7 & 30.3→82.1 \\
    \midrule
    LD2 & Neg. & Top 10\% & mark carney's economic policies & 15.0→3.4 & 50.0→16.1 & 35.0→80.5 \\
     &  &  & trump & 9.8→25.4 & 35.2→22.3 & 55.0→52.3 \\
     &  &  & donald trump & 19.6→42.0 & 14.9→3.8 & 65.5→54.1 \\
    \midrule
    LD2 & Neg. & Top 1\% & liberal party policies & 50.0→2.8 & 0.0→2.8 & 50.0→94.4 \\
     &  &  & liberal government approach & 50.0→2.8 & 0.0→2.8 & 50.0→94.4 \\
     &  &  & indigenous immigration policies & 37.8→10.5 & 18.9→4.7 & 43.2→84.9 \\
    \bottomrule
    \end{tabular}
    \caption{\rev{Examples of significant stance change on stance targets that are loaded on by the first two latent dimensions, for Instagram. Numbers/arrows indicate old and new percentage over time.}}
    \label{tab:instagram-examples}
\end{table*}

In Table \ref{tab:bluesky-examples} we include the top three significant ($p < 0.05$) examples of stance changes on targets for Bluesky from our dataset.

\begin{table*}[]
    \centering
    \begin{tabular}{llllrrr}
    \toprule
    Dim. & Dir. & Percentile & Target & Favor & Neutral & Against \\
    \midrule
    LD1 & Pos. & Top 10\% & indigenous rights & 64.5→19.9 & 34.6→80.1 & 0.9→0.0 \\
     &  &  & economy & 67.6→31.7 & 16.2→41.8 & 16.2→26.5 \\
     &  &  & ndp & 61.9→35.4 & 38.1→43.8 & 0.0→20.8 \\
    \midrule
    LD1 & Pos. & Top 1\% & climate change policy & 59.0→25.7 & 38.5→45.7 & 2.6→28.6 \\
     &  &  & fossil fuel phase-out & 61.5→37.1 & 38.5→37.1 & 0.0→25.7 \\
     &  &  & carbon emissions regulation & 48.7→34.3 & 48.7→45.7 & 2.6→20.0 \\
    \midrule
    LD1 & Neg. & Top 10\% & covid vaccine policy & 39.0→67.5 & 58.1→31.2 & 2.9→1.3 \\
     &  &  & covid vaccine rollout & 43.4→72.7 & 55.1→26.0 & 1.5→1.3 \\
     &  &  & economy & 46.7→71.1 & 15.6→20.0 & 37.8→8.9 \\
    \midrule
    LD2 & Pos. & Top 10\% & donald trump & 36.4→8.8 & 18.2→11.0 & 45.5→80.2 \\
     &  &  & president & 26.7→7.5 & 53.3→22.5 & 20.0→70.0 \\
     &  &  & american diplomacy & 5.4→3.5 & 75.5→65.7 & 19.1→30.8 \\
    \midrule
    LD2 & Neg. & Top 10\% & elon musk & 1.7→7.0 & 25.4→59.6 & 72.9→33.3 \\
     &  &  & trump & 5.8→4.4 & 26.9→41.8 & 67.3→53.8 \\
     &  &  & mark carney & 42.0→12.8 & 34.0→45.9 & 24.0→41.3 \\
    \bottomrule
    \end{tabular}
    \caption{\rev{Examples of significant stance change on stance targets that are loaded on by the first two latent dimensions, for Bluesky. Numbers/arrows indicate old and new percentage over time.}}
    \label{tab:bluesky-examples}
\end{table*}

\end{document}